# A Systematic Mapping Study on Smart Cities Modeling Approaches

MARIA TERESA ROSSI, University of Milano-Bicocca
MARTINA DE SANCTIS, Gran Sasso Science Institute, Italy
LUDOVICO IOVINO, Gran Sasso Science Institute, Italy
MANUEL WIMMER, Johannes Kepler University, Austria

The Smart City concept was introduced more than one decade ago to define an idealized city characterized by automation and connection. It then evolved rapidly by including further aspects, such as economy, environment, citizens, to name just a few. Since then, many publications have explored various aspects of Smart Cities across different application domains and research communities, acknowledging the interdisciplinary nature of this subject. In particular, our interest focuses on how smart cities are designed and modeled, as a whole or as regards with their subsystems, when dealing with the accomplishment of the research goals in this complex and heterogeneous domain. To this aim, we performed a systematic mapping study on smart cities modeling approaches identifying the relevant contributions ($i$) to get an overview of existing research approaches, ($ii$) to identify whether there are any publication trends, and ($iii$) to identify possible future research directions. We followed the guidelines for conducting systematic mapping studies by Petersen *et al.* to analyze smart cities modeling publications. Our analysis revealed the following main findings: ($i$) smart governance is the most investigated and modeled smart city dimension; ($ii$) the most used modeling approaches are business, architectural, and ontological modeling approaches, spanning multiple application fields; ($iii$) the great majority of existing technologies for modeling smart cities are not yet proven in operational environments; ($iv$) diverse research communities publish their results in a multitude of different venues which further motivates the presented literature study. Researchers can use our results for better understanding the state-of-the-art in modeling smart cities, and as a foundation for further analysis of specific approaches about smart cities modeling. Lastly, we also discuss the impact of our analysis for the Model-Driven Engineering community.

## 1 INTRODUCTION

The concept of Smart City (SC) was introduced in the United States as an IBM's brand in 2011 [31] to define an idealized city characterized by automation and connection. It further evolved detailing the multiple aspects characterizing SCs themselves, which must be considered to effectively make cities smarter, such as sustainable mobility, environmental management, citizens inclusion, and so on. Today, the concept of SC is still evolving, and there is not a universally accepted definition recognized by the research communities. Thereafter, in 2014 Boyd Cohen provided the so called Boyd Cohen Wheel [12], representing the commonly used standard in the literature to describe the constituent dimensions

Authors' addresses: Maria Teresa Rossi, maria.rossi@unimib.it, University of Milano-Bicocca, Milan, Italy; Martina De Sanctis, martina.desanctis@gssi.it, Gran Sasso Science Institute, L'Aquila, Italy; Ludovico Iovino, ludovico.iovino@gssi.it, Gran Sasso Science Institute, L'Aquila, Italy; Manuel Wimmer, manuel.wimmer@jku.at, Johannes Kepler University, Linz, Austria.

of SCs, namely Smart Mobility, Smart Environment, Smart Economy, Smart People, Smart Living, Smart Governance. In this work, we further extend it to include Smart Things, to also consider the relevant role played by IoT networks in SCs. Since then, the concept of SC has been gaining more and more relevance both at European and world-wide level. This is because weaving Information and Communication Technology (ICT) and cities can be exploited in order to promote their sustainable development to improve human lives and preserve the environment. For instance, the European Commission defined a list of Sustainable Development Goals (SDGs)[1] to guide cities development from this perspective. Indeed, the smart development of a city is of interest of different research communities and practitioners, for instance computer scientists, engineers, social scientists, economists, but also public/private companies, public administrations, urban experts.

Despite the transition of cities to SCs can lead to a lot of benefits at a sustainability and inclusiveness level, this type of development is difficult due to the complex nature of cities. In fact, as we have seen, multiple dimensions (e.g., mobility, economy) come into play, together with their corresponding stakeholders (e.g., private companies, public administrations). These dimensions are very heterogeneous and this makes the interconnection among them difficult. Moreover, the transition to SCs might also be affected by the growing (and uneven w.r.t. the different dimensions) amount of investments in smart initiatives and the advent of new technologies. As a consequence of these high complexity and heterogeneity, SC experts coming from different research communities give rise to a multitude of different models used to design and describe SCs. This proliferation of SC models could be counter-productive by further hindering the interconnection among the diverse dimensions, thus the communication among their stakeholders.

In the field of software and system modeling research, there have been efforts to address the challenge of defining multiple and heterogeneous SC models. For instance, in [33] the authors present a Domain-Specific Language (DSL) to model Smart City Systems (SCSs). Here, the authors face the high heterogeneity of devices and communication protocols usually adopted while creating a SCS. Instead, in [2] the author exploits Model-Driven Engineering (MDE) to afford the issue related to the collaboration between the different stakeholders in the SC domain. They provide a reference architecture for SC projects presented as a design language for creating SCs blueprints useful to model stakeholders and technologies coming from a SC.

Considering that SCs are under discussion for about ten years, various papers capturing different aspects of this topic have been published at different venues by different research communities. Moreover, given the multi-disciplinary nature of SCs, there are many application fields where they have been applied. The high diversity of the involved disciplines, the many dimensions making SCs (as reported in the Boyd Cohen Wheel [12]) and the need for designing and further integrating them, clearly highlight the relevance of *modeling* in the domain of SCs, and the need for understanding the role that model-driven engineering can play and where we are so far. Furthermore, the lack of modeling standards for SCs and the unclear applicability of existing modeling approaches, combined with new emerging technologies, such as digital twins of cities (e.g., urban digital twins [14]), which also require a solid model base, make it urgent to investigate the state of the art about SCs modeling approaches. In this context, to better understand how SCs have been modeled from the perspective of the different research communities and in the context of diverse application fields, we performed a systematic mapping study. It aims to collect and analyse the existing SC modeling approaches in such a way to provide an overview about the state of the art on SC modeling. In particular, we performed specific analysis to identify possibly emerging SC dimensions, application fields, modeling approaches, publication venues, and to determine existing trends, if any, and potential future directions that SCs modeling should take. We further investigate the maturity level of the analyzed works to understand to which extent the provided solutions are technically sound and robust. We believe that such analyses might be quite helpful for the multitude of researchers and practitioners actively working on this topic, by expanding their knowledge on this domain and identifying future directions not yet widely investigated.

The remainder of this article is structured as follows. Section 2 discusses related surveys about the design of SCs. In Section 3, an overview of the research methodology of our mapping study is given. Section 4 analyzes the results of our study w.r.t. the stated research questions. Section 5 summarizes the research findings and gives some future directions. Lastly, in Section 6 the threats to validity of our study are discussed, whereas Section 7 concludes the work.

## 2 RELATED WORKS

In this section, we report related surveys in which the SC domain and its sub-domains have been inspected. In particular, we focus on those surveys whose goal is to investigate the *design* of SCs. We grouped the reviewed surveys according to the topic in the SC domain they relate to.

**SC Concepts.** Literature documents a great effort in defining the concept of SC in different research communities such as software engineering and social sciences. For instance, Sanchez-Corcuera *et al.* [37] report a survey on investigating the SC domain through its various definitions, applications, and challenges. The aim of the survey is to emphasize the opportunities in relation to the challenges that need to be addressed in the domain, with the goal of enhancing the level of smartness within cities. For instance, the implementation of a smart traffic monitoring system offers significant advantages, enabling real-time data collection for efficient traffic management. This technology can enhance urban mobility, decrease congestion, and improve residents' quality of life. However, challenges within the public administration (e.g., slow decision-making processes, limited funding) can hinder the system's implementation. Additionally, concerns regarding privacy may restrict data collection and usage, reducing the system's potential effectiveness. Achmad *et al.* [3] performed a literature review on existing SCs conceptual models and frameworks to analyze their dimensions, working areas, and characteristics. Thus, they propose a classification of SCs resources (e.g., network infrastructure, data centers) and services (e.g., environmental governance, investments management) in order to help future developments of SCs models and frameworks. In contrast, Mishbah *et al.* [25] present a systematic review to investigate the smart village concept and extract objectives, strategies, dimensions, and foundations of this specific domain. The concept of smart village is presented as a series of initiatives whose goal is related with the development of rural areas. The authors' goal is that of designing a conceptual model useful to represent smart villages to support projects implementation. Anthopoulos *et al.* [6] compare modelling and benchmarking approaches used to conceptualize and evaluate SCs. Interestingly, the authors highlight that the compared approaches have in common six dimensions, namely people, government, economy, mobility, environment and living. These dimensions match those previously defined by Boyd Cohen in [12] and are widely used in the SCs domain. They eventually use existing analysis models for SCs on top of three representative SCs as use cases.

**SC Data Management.** This aspect gained a lot of attention in research. This is due to the challenges related to the collection, integration, and manipulation of the increasing volume of urban data produced by SCs. Moustaka *et al.* [26] presented a systematic literature review investigating the core topics (e.g., IoT, crowd sensing), services, and methods applied in SCs data monitoring. The review results are presented by means of taxonomies describing the state-of-the-art of the analyzed primary studies. The objective of the review is to propose guidelines for the use of data science techniques in the SC domain. The survey proposed by Perera *et al.* [28] also belongs to the data analytics context. The motivations of the survey are related to the issues coming from the availability of huge amounts of data generated by a variety of smart objects. Thus, the authors investigate the Fog computing domain w.r.t. the SC sustainability.

**SC Governance.** Models to support decision-making processes gained a lot of attention in the *smart governance* dimension. Pereira *et al.* [27] focus their literature review on the smart governance. In particular, they investigate the use of ICT in this context to improve decision-making through better collaboration among the different stakeholders, including government and citizens. Moreover, they extract smart governance models addressing challenges such as information sharing, citizen engagement, transparency, and openness. Ma *et al.* [23] investigated modeling and decision making methodologies associated with the utilization of smart services in a city. In their survey, they extract the key issues related to SCs data and decision-making (e.g., heterogeneity, privacy, interdisciplinarity). The survey is motivated from the fact that the available large amount of city-generated data across multiple domains is not fully exploited. Meanwhile, Giang *et al.* [16] propose a review of methodologies for modelling SC Living Labs, i.e., the collaboration in decision-making processes involving multiple stakeholders. The authors present the concept of SC as a set of complex and dynamic networks that, through collaboration and interaction, have to achieve common goals and make a final decision. Rodríguez [30] analyses models of citizen participation and the use of new technologies by city governments in SCs in order to improve e-participation of the citizens. He highlights the need for using new technologies to support the adoption of a more participative model of governance. In smart governance there is also the need of finding methods to evaluate SCs in order to create rankings. Benamrou *et al.* [9] present a comparative study among three models (i.e., Ruddolf Giffenjer European ranking of medium-size SCs [17], Mapping SCs in the European Union [24] and Boyd Cohen wheel models [12]) used to evaluate and classify SCs. They compare different aspects (e.g., data sources, calculated indicators) to find the most suitable model for Maroccan SCs. Kulatunga *et al.* [21] propose a systematic literature review on Performance Measurement models used to measure the smartness of cities, i.e., the performance. The review investigates the different available definitions for smartness of a SC. The authors find that, rather than an exact definition, there are several common characteristics/indicators used by SCs frameworks to measure performances.

**SC Mobility and Tourism.** Much attention is also addressed to the *smart mobility* and *smart tourism* dimensions of SCs. For instance, Solmaz *et al.* [41] survey the currently used human mobility models with a particular focus on the commonly used metrics and data collection techniques. They investigated this field since it represents a key component of various research areas including transportation, mobile networks, disaster management, urban planning, and epidemic modeling. Furthermore, Chowdary *et al.* [11] introduce the concept of Internet of Vehicles (IoV) as an important part of IoT when the prime focus is on vehicles as the underneath intelligent devices. The authors evaluate different mobility models highlighting their influence on different vehicular network performance parameters. In the smart tourism context, Kontogianni *et al.* [20] investigate social network user modeling in their literature review. In particular, they focus on smart tourism applications and user modeling via social network data analysis in order to extract useful information to support tourism services development.

**Other SC Topics.** Other surveys put their attention on articles exploiting specific modeling approaches in the SC domain. For instance, Souza *et al.* [13] made a literature review in which they investigate papers discussing sustainability issues linked with business models and business model innovation in the IoT context. Their goal is to raise the major issues in the IoT context related to sustainability, in order to help researchers in developing new solutions. Shetty *et al.* [39] investigate through a systematic literature review various business models developed for the SCs domain, in order to understand if by using this type of modeling approaches it is possible to discover the economic value generated by a SC. Aljowder *et al.* [4] investigate the research progress in SC maturity models through a systematic literature review. They analyze some maturity frameworks focusing also on the alignment of the SC to the United Nations Sustainable Development Goals (SDGs[2]) in order to promote smartness transformations for cities. Torrinha *et al.* [42]

---

[2] https://sdgs.un.org/goals

| Survey Paper | Year | Application Domain | Survey's Subject(s) |
|---|---|---|---|
| Anthopoulos *et al.* [6] | 2015 | Smart City | Conceptual Models; Benchmark Tools |
| Benamrou *et al.* [9] | 2016 | Smart Governance | Ranking Models |
| Chowdary *et al.* [11] | 2016 | Smart Mobility | Mobility Models |
| Perera *et al.* [28] | 2017 | IoT | Fog (Edge) Computing |
| Giang *et al.* [16] | 2017 | Smart Governance | Smart City Living Lab projects |
| Torrinha *et al.* [42] | 2017 | Smart Governance | Maturity Models |
| Achmad *et al.* [3] | 2018 | Smart City | Conceptual Models |
| Mishbah *et al.* [25] | 2018 | Smart Village | Conceptual Models |
| Moustaka *et al.* [26] | 2018 | Data Management | Data Harvesting and Data Mining Processes |
| Rodríguez [30] | 2018 | Smart Governance | Collaborative Models; Participative Models |
| Kontogianni *et al.* [20] | 2018 | Smart Tourism | User Models |
| Pereira *et al.* [27] | 2018 | Smart Governance | Smart Governance Definitions |
| Lim *et al.* [22] | 2018 | Urban Planning | Reference Models |
| Sanchez-Corcuera *et al.* [37] | 2019 | Smart City | Smart City Definitions |
| Ma *et al.* [23] | 2019 | Smart Governance | Data Management; Services Models; Decision-making |
| Solmaz *et al.* [41] | 2019 | Smart Mobility | Mobility Models |
| Souza *et al.* [13] | 2019 | IoT | Business Models |
| Shetty *et al.* [39] | 2019 | Smart Economy | Business Models |
| Aljowder *et al.* [4] | 2019 | Smart Governance | Maturity Models |
| Abbasi *et al.* [1] | 2019 | IoT | Conceptual Models |
| Kulatunga *et al.* [21] | 2020 | Smart City | Performance Measurement Models |
| **Our Paper** | | **Smart City** | **Modeling Approaches** |

Table 1. Comparison of related surveys.

identify SC maturity models and assess them w.r.t. their current relevance. Their objective was that of highlighting the need of completeness when designing maturity models, given that, in the SC domain, they are used to support investment decisions. Abbasi *et al.* [1] examine various applications of the existing conceptual modeling approaches for IoT. Firstly, they map conceptual modeling approaches on the different layers of the functional model of the IoT reference architecture [8]. In addition, they show the use of different modeling approaches organized as a hierarchical model for a complex IoT system. Lim *et al.* [22] propose four reference models to support the extraction of information from big data in real-world applications. In particular, they focus on the transformation of urban data in such information models to support urban planning.

**Synopsis.** We report in Table 1 all the analyzed related work to give an overview of the recent surveys in the SCs domain. The table reports also a classification of the surveys w.r.t. the investigated domain and survey's subject(s). Specifically, given the objective of our study, the majority of the surveyed subjects correspond to modeling approaches. However, also processes, tools and other techniques have been considered. Interestingly, the majority of works either focus on a specific SC dimension (e.g., Smart Governance, Smart Mobility, Smart Economy) or on one or a few particular subjects (e.g., Conceptual Models, Business Models, Maturity Models). Only a few of them (i.e., [6], [3], [37], [21]) investigate the SC domain in its entirety. However, these surveys are focused only on specific subjects, as highlighted in Table 1. Differently, our study aims to investigates the entire SCs domain but without looking for particular subjects. On the contrary, we aim to synthesize from the mapping study which modeling approaches are used in the analyzed domain, which of them are used the most, and in which sub-domains of the SCs domain they are applied.

## 3 RESEARCH METHODOLOGY

To conduct our research investigation about the modeling approches for SCs, we performed a systematic mapping study (SMS) with the aim of screening and classifying publications in the given area. In particular, we followed the process for systematic mapping studies in software engineering proposed by Petersen *et al.* [29]. We have adhered to the reported guidelines, by carrying out all the given process steps, with some adjustments in the classification and mapping phases, due to further investigations we were interested in, as we detail in the rest of this section.

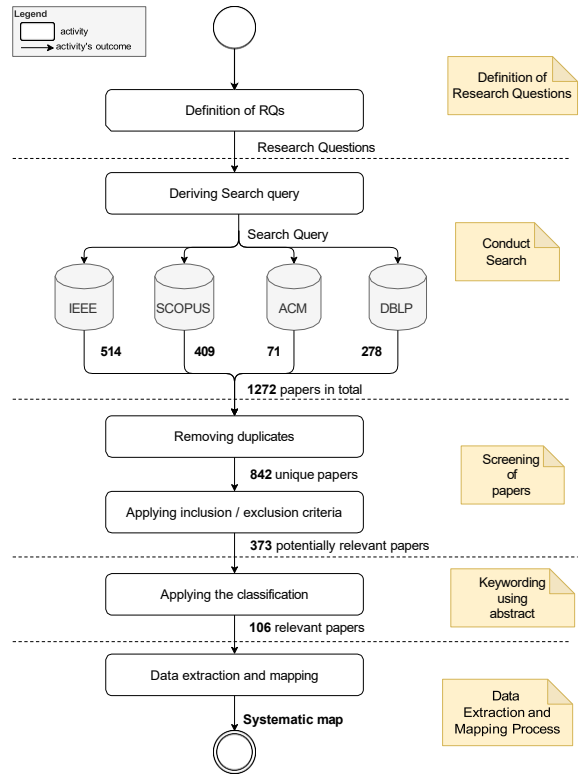

Fig. 1. The used systematic mapping process based on the guidelines of Petersen *et al.* [29].

The process we followed for accomplishing the mapping study is depicted in Figure 1. Yellow notes on the right hand side refer to the corresponding process steps defined by Petersen *et al.* [29]. The main part of Figure 1 depicts our systematic mapping process, by highlighting both the performed activities and corresponding outcomes. The starting activity consists in the definition of research questions. As outcome, we gathered a set of research questions delineating the review scope of this study and focusing on the modeling of SCs as their main topic (see Section 3.1). The search query is derived from the stated research questions and used to conduct a literature search on the selected digital libraries (see Section 3.2). In particular, libraries were surveyed in October 2021 and our focus is on the publications published until 2020 included, without specifying any lower bound to the time range. The outcome of this search activity consists in all the publications related to the topic, namely **1272** papers. It is then followed by a screening of these publications aiming at removing any duplicates, in addition to those publications that, according to the inclusion/exclusion criteria, are not relevant to answer the research questions (see Section 3.3). This allowed us to get a set of **373** unique papers representing the input for the classification step. The classification is the most elaborate step, together with the final mapping. It essentially involves concurrently performing keywording on publication abstracts and categorizing publications based on various aspects (see Section 3.4). Specifically, in order to find out the maturity level of the analyzed studies, we categorized the relevant publications according to the so-called *research type facets* (see Table 2), introduced by Petersen *et al.* [29]. Furthermore, we augmented the classification activity by performing further analysis due to our interest in investigating other aspects of the relevant studies. As a consequence, we categorized publications also according to (*i*) the *Technology Readiness Level* (TRL) scale[3] (see Table 3), (*ii*) the *smart city dimensions*, based on the Boyd Cohen Wheel standard [12], (*iii*) the *modelling approaches* exploited or proposed as contribution in the publications, (*iv*) the presence of a specific implemented *tool* based on the presented contribution, i.e., web applications, prototype, services, as outcome of the studies, and (*v*) the *application field of the contribution* in the reference domain. These additional analysis asked for a deeper look at the papers, that goes beyond the title and abstract, which allowed us to identify additional papers whose contribution was out of the scope of this study. Consequently, as output of the classification step we obtained a set of **106** relevant papers, namely those used for the final mapping activity (detailed in Section 4) resulting in the systematic map.

In the rest of this section we describe each process step, apart for the final mapping of papers that is reported in Section 4 where the main findings of this work are extracted and discussed.

All the data, such as papers dataset, performed classifications and results, and used tools can be found in the online replication package [34].

### 3.1  Stage 1: Identification of the Research Questions

In order to determine the review scope of this mapping study, we started by defining our research questions (RQs). They reflect the objectives of this study, namely identifying the research status about modelling approaches for SCs, either as a whole or by considering their multiple variations and dimensions, the maturity level of these approaches, and the open challenges that still need to be addressed. In addition, we are interested in discovering if it exists a model for SCs that is recognized as a benchmark for one or multiple scientific community. We grouped the RQs with respect to the domain they belong to, namely problem, solution and scientific community domains.

— **Problem Domain:** concerning the problem domain, we are interested in understanding which SC challenges are addressed and in which SC dimensions by using modeling approaches.

**RQ1** *Which Smart City dimensions have been modeled the most?*

The intention behind this RQ is to identify for which SCs dimensions, namely Smart Mobility, Smart Environment, Smart Economy, Smart People, Smart Living, Smart Governance, Smart Things, a uniform way of modeling their concepts and relationships has been used. Furthermore, we are interested in identifying which SCs dimensions attracted more the interest of the research communities, if any.

**RQ2** *Which application fields in the Smart City domain have been modeled the most?*

The aim of this RQ is to find out how SCs modeling approaches can help w.r.t. the management of the common SCs issues. Replying to this RQ might also help understanding the challenges which have been addressed the most.

— **Solution Domain:** concerning the solution domain, we are interested in knowing about the modeling approaches currently used to design SCs as well as SCs subsystems, and their status in terms of maturity.

---

[3]  https://enspire.science/trl-scale-horizon-europe-erc-explained/

**RQ3** *Which modeling approaches have been used to represent Smart Cities?*

The aim of this RQ is to investigate the different modeling approaches and techniques that have been used to model SCs and/or their dimensions.

**RQ4** *What is the maturity status of smart cities modeling approaches?*

The intention behind this RQ is that of finding out the maturity level of the analyzed works. This might also help in understanding in which directions a major research effort is required.

— **Scientific Community Domain:** with respect to the scientific community domain, we are interested in understanding from how long modeling approaches started to be used and applied in the SC domain and which venues (i.e., conferences, workshops and journals), are targeted the most to disseminate the results.

**RQ5** *When did the contributions on modeling Smart Cities occur?*

The intention behind this RQ is to know when modeling started contributing to SCs, what has been the trend from its beginning up to 2020, and the type of the surveyed publications (e.g., journal articles, conference papers).

**RQ6** *Where have the contributions been published?*

Throught this RQ we are interested in catching out the most relevant venues and journals where papers on SCs modeling have been published.

### 3.2   Stage 2: Conducting the Search

The identification of primary studies was driven by the defined RQs and the identified keywords that were used, in turn, to formulate the search query to apply in digital libraries. Differently from the existing works listed in Table 1, we are not interested in a single or a few modeling approaches or model types exploited in the domain of SCs (e.g., maturity models, business models, UML). On the contrary, we are interested in observing the trends of the used modeling approaches in SCs as they emerge from the search, thus to get an overview on the domain from the surveyed papers. For these reasons, we agreed about the use of the keywords *smart city* and *model* that identify the domain of the study and the investigated approaches, respectively. These keywords and their synonyms (e.g., smart cities, modelling, modeling) have been used to shape the schema of the search query reported in (1):

$$(\textit{smart city} \textbf{ OR } \textit{smart cities}) \textbf{ AND } (\textit{model} \textbf{ OR } \textit{modeling} \textbf{ OR } \textit{modelling}) \tag{1}$$

Among the established digital libraries available online, we selected the following four to run the automated search, to guarantee a certain level of quality of relevant papers:

(1) *IEEE Xplore*[4];

(2) *Scopus*[5];

(3) *Association for Computing Machinery (ACM)*[6];

(4) *dblp computer science bibliography*[7].

For each library, we submitted the query in (1), with the appropriate syntax, as reported below:

· **IEEE** ("Document Title": "smart city" OR "smart cities") AND ("Document Title": model);

· **Scopus** TITLE ( "smart city" OR "smart cities" ) AND TITLE ( model );

· **ACM** [[Publication Title: "smart city"] OR [Publication Title: "smart cities"]] AND [Publication Title: model];

· **dblp** smart cit* model*.

---

In order to make an accurate search, thus getting more precise results effectively focused on *the use of modeling approaches in the SCs domain*, we applied a restriction to the search query. Specifically, we considered only those publications having the stated keywords in the title. This way, given that the used keywords have a broad meaning subject to multiple interpretations, we could discard those papers mentioning SCs and modeling in the abstract and / or keywords without being effectively focused on this study's topic. Indeed, without this restriction and by considering also abstract and keywords in the search, we were passing from hundreds to thousands publications. We are aware that publication bias are possible, as in any systematic study, and we discuss this aspect in the threats to validity section (see Section 6). Although this restriction is easy to apply in the IEEE Xplore, Scopus and ACM libraries, as shown by the queries above, it is not possible in the dblp library. In dblp, indeed, one can only specify if the search refers to authors, venues, publications, or a combination of them, and a publications-based search looks also into other fields (e.g., venue, publisher). As a consequence, the query on the dblp library returned several out of scope publications (e.g., with the searched keywords appearing in the publications' venue name instead than in their titles), which however have been identified and discarded by means of the title analysis and keywording performed during the publications screening and classification steps.

### 3.3   Stage 3: Screening of Publications

The screening of publications started by removing the duplicates, which easily occur when querying multiple libraries. Furthermore, we defined a set of *Inclusion / Exclusion Criteria* to identify the set of potentially relevant papers to be taken into account. The selection criteria that we defined are listed below:

- Inclusion Criteria:
  - Publication year $\leq$ 2020;
  - Peer-reviewed studies published in journals, conferences, and workshops;
  - Studies available for consultation.
- Exclusion Criteria:
  - Papers not available in English;
  - Studies of less than or equal to two pages, e.g., calls for papers, slides;
  - Book chapters, since they might be not peer-reviewed;
  - Mapping studies or systematic literature reviews on the topic that are not offering a contribution, i.e., in the form of a taxonomy or conceptual model for SCs;
  - Out of scope papers;
  - Papers not providing enough information about the modeling approach.

In particular, the last two exclusion criteria have been further applied during the stage 4, namely keywording using abstracts, where multiple out of scope publications as well as publications *not contributing with any kind of modeling approach* were identified. This lead the set of relevant papers to pass from 373 to 106. We are aware that our search strategy might look overly restrictive, however we argue that the discarded papers did not contribute a modelling approach, not even a conceptual one. This can be verified in the replication package [34].

### 3.4   Stage 4: Keywording using abstracts

The keywording using abstracts served multiple objectives. On the one hand, it allowed to detect further out of scope papers, by means of an *analysis* focused on title, abstract, and keywords. During this assessment phase, we considered those papers focusing on developing or applying modeling approaches, specifically in the realm of model-driven engineering, in the SC domain, while we excluded those papers dealing with modeling approaches out of our scope (e.g., machine learning, social network analysis). On the other hand, it supported the classification of publications. Indeed, during this phase, we categorized the relevant papers according to multiple aspects, also through a deeper analysis (i.e., an overview to the entire body of the publication). We started by identifying the *research type facets*, reported by Petersen *et al.* [29], and previously introduced by Wieringa *et al.* [44]. Facets are listed in Table 2, and essentially allow the classification of papers in two main categories, namely non-empirical papers (i.e., experience, opinion, philosophical and solution proposals), and empirical papers (i.e., validation and evaluation of solutions) reporting about research evaluated in the context of real world projects, in cooperation with industry or showing experiments or statistics.

Table 2.  Research type facets from Petersen *et al.* [29]

| Category | Description |
| --- | --- |
| Experience Papers | Experience papers explain on what and how something has been achieved in practice. It has to be the personal experience of the author. |
| Opinion Papers | Opinion papers express the personal opinion of somebody whether a certain technique is good or bad, or how things should be done. They do not rely on related work and research methodologies. |
| Philosophical Papers | These papers sketch a new way of looking at existing things by structuring the field in form of a taxonomy or conceptual framework. |
| Solution Proposal | A solution for a problem is proposed. The solution can be either novel or a significant extension of an existing technique. The potential benefits and the applicability of the solution is shown by a small example or a good line of argumentation. |
| Validation Research | The investigated solutions are novel and have not yet been implemented in practice. Solutions used are for example experiments, i.e., work done in the lab. |
| Evaluation Research | Solutions are implemented in practice and an evaluation of the solutions is conducted. That means, it is shown how the solution is implemented in practice (solution implementation) and what are the consequences of the implementation in terms of benefits and drawbacks (implementation evaluation). This also includes to identify problems in industry. |

Moreover, in order to find out the maturity level of the analyzed works, we measured the *Technology Readiness Level* (TRL) of those publications belonging to the solution, validation and evaluation facets. The TRL scale [15] we exploit, is the one used by the EU funded projects arena that, in turn, was defined by NASA in the 1990's as a means for measuring the maturity of a given technology. Without loss of generality, we used it to measure the maturity level of the modeling approaches reported in the analysed works. The TRL scale we referred to is reported in Table 3, while the tool used for the calculation in our replication package[34].

Table 3.  Technology Readiness Level scale [15].

| Level | Description |
| --- | --- |
| TRL 1 | Basic principles observed |
| TRL 2 | Technology concept formulated |
| TRL 3 | Experimental proof of concept |
| TRL 4 | Technology validated in lab |
| TRL 5 | Technology validated in relevant environment |
| TRL 6 | Technology demonstrated in relevant environment |
| TRL 7 | System prototype demonstration in operational environment |
| TRL 8 | System complete and qualified |
| TRL 9 | Actual system proven in operational environment |

Additionally, during the keywording stage, we performed further analysis to enrich and drive the classification of papers with relevant information for the considered domain of the study. Specifically, we categorized papers according to:

(1) The *smart city dimensions*, based on the Boyd Cohen Wheel standard [12]. From this classification, we expect to derive which dimensions have benefited the most from the application of one or more modeling approaches. We aim to understand also which dimensions show the lack of exploitation of modeling approaches, which might highlight open challenges requiring for future research efforts.

(2) The *modelling approaches* exploited or proposed as contribution in the publications. Through this classification we aim to understand which modeling approaches and/or languages are most often used.

(3) The presence of a specific *implemented tool* based on the presented contribution, i.e., web applications, prototype, services, as outcome of the studies. This analysis contributes also in deriving the maturity level of the surveyed proposals, which is particularly relevant if we consider that solutions in the SCs domain are often addressed to stakeholders different than researchers, such as municipalities employees, to give just one example.

(4) The *application field* of the contribution in the reference domain. This classification is relevant to understand which domains are targeted the most as well as those less covered, among the multiple domains under the SCs sphere.

This multi-facet classification allowed us to derive the systematic mapping reported in the next section.

## 4  MAPPING OF PAPERS

After the extraction and definition of the final set of relevant papers, we performed the mapping of papers. This process was possible thanks to keywording performed in the previous classification phase. In this Section we report the analysis that we did on the final set of publications w.r.t. the defined RQs and by exploiting the categories described in Section 3.4.

### 4.1  RQ1: Which Smart City dimensions have been modeled the most?

To answer RQ1 we assigned each publication to one of the SC dimensions defined in the Boyd Cohen Wheel standard [12]. We selected it among other works about SC dimensions (e.g., [6], [18]), since the Boyd Cohen Wheel represents the commonly used standard in the literature to describe the constituent dimensions of SCs. In Figure 2 we report a revised version of the Boyd Cohen Wheel where we added a further dimension specifically for Smart Things. We made this extension to properly classify (and keep track of) those publications whose contribution lies mainly in the IoT world, albeit they might belong to one of the other dimension. Indeed, we highlight here that, as refer to the performed classification, we are aware that a paper might belong to more than one dimension. In this case, we selected the most fitting one. Eventually, we have seven dimensions [40], which we briefly describe in the following:

· *Smart Economy:* it is characterized by the exploitation of ICT in economic activities. In other words, it refers to those activities aiming at creating an economic value for the city. For this reason, we included in this dimension also those publications dealing with *smart tourism*, due to its economic implication [19].

· *Smart Environment:* it concerns the control and monitoring of environmental factors, such as pollution, planning of green areas, waste. Initiatives in this dimension target an effective and efficient use of public natural resources means (e.g., alternative energy sources) to avoid fossil fuels, to reduce carbon footprint, etc.

· *Smart Governance:* it concerns the use of technology to enable open, transparent and participatory governments. The contributions in this dimension are mainly focused in supporting decision-making processes for cities governments.

· *Smart Living:* it concerns the improvement of the quality of people's daily life and lifestyle. It mainly regards contributions about smart buildings, i.e., systems with appliances and services connected in a building's network (e.g., homes, hospitals), with the aim of optimizing resources consumption and management.

· *Smart Mobility:* it has the aim of improving local accessibility to mobility services and a smart and sustainable mobility, while also supporting social inclusion, reducing the environmental impact, etc. As part of this dimension we mainly find contributions about the development of sustainable, innovative and safe transport systems and applications.

· *Smart People:* it deals with the promotion of creativity, open-mindedness and participation in public life. This dimension includes contributions about services for citizens (e.g., e-learning platforms, public administration online services), investment on and preservation of the human capital, and privacy-related solutions (e.g., privacy protection systems) [7].

· *Smart Things:* it refers to the ever-growing network of physical objects equipped with an IP address for Internet connectivity, and the communication between these objects as well as with other Internet-enabled devices and systems [32].

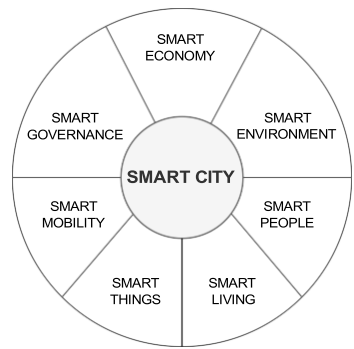

Fig. 2.  Boyd Cohen Wheel [12] extended with the Smart Things dimension.

We recall here that in this mapping study we considered those publications whose contribution includes one or more modeling approaches and targets the SCs as application domain. In Figure 3, we reported the number of publications per SC dimension. We can see that the majority of publications fall in the smart governance dimension (**37**). The second most prominent dimension is the smart environment (**23**), while the smart people follows at the third place of the classification (**13**). Apart the smart governance and smart environment, the distribution of publications among the other dimensions is quite uniform, with the smart things closing the classification (**5**).

Moreover, we made a further analysis on the data extracted to reply to RQ1 and we looked at the distribution of publications per year w.r.t. the SC dimensions. Results are show in Figure 4. This analysis gives an overview on which dimensions gained more attention over the years and when corresponding publications started to appear. We can observe that the smart governance, smart environment and, to some extent, the smart people dimensions follow a constant distribution, after their first appearance. This indicates that the interest in this fields, despite varying, remained constant over the years. As regards the publications fitting in the remaining dimensions, they follow a quite fluctuating distribution. Moreover, some of them, as for instance the smart things, smart mobility and smart living publications started appearing later on, in 2014, 2015 and 2016 respectively. This might be because the investments in IoT technologies, widely used also in smart mobility and living approaches (e.g., smart traffic lights, smart buildings), came after the advent of the SC concept in 2011.

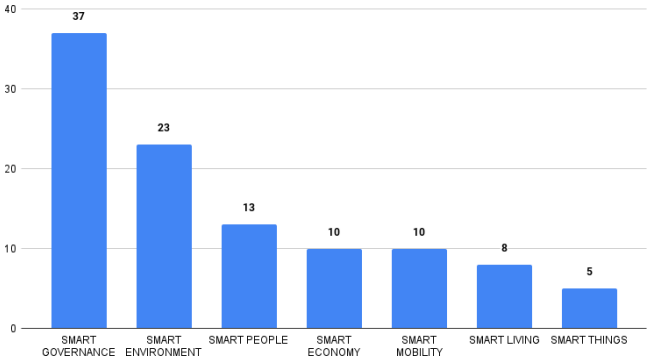

Fig. 3. Distribution of publications among the smart city dimensions.

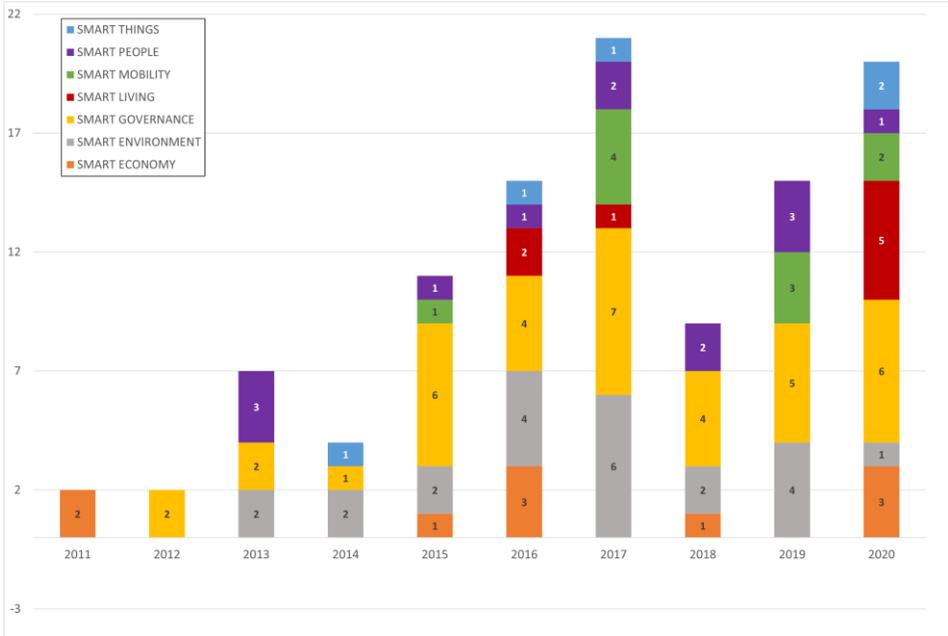

Fig. 4. Number of publications per year w.r.t. the smart city dimensions.

### 4.2   RQ2: Which application fields in the Smart City domain have been modeled the most?

The application fields have been extracted during the keywording phase. Specifically, we let them emerge from the analyzed publications. To do so we followed a systematic process where we first extracted the emerging application fields. Then, we grouped them by synonyms to make uniform application fields' classes. Lastly, we organized them hierarchically based on the relations among classes (e.g., Sustainability class under the Quality class). The process involved several iterations among authors. We arranged the identified application fields in the taxonomy that we reported in Fig. 5. This schema gives an overview on which are the application fields of SCs modeling approaches as the literature evidence provides. The taxonomy is organized in different hierarchical levels depicted with different colors, going from the more generic to the more specific ones, by navigating it from the left to the right. The numbers in brackets reported in each class of the taxonomy represent the number of papers that fall into that class. We specify here that the numbers reported for inner classes (i.e., the blue and green classes in Fig. 5) might differ from the sum of the papers counted in their sub-classes, since inner classes also contain those papers which are more generic w.r.t. the fields identified by the sub-classes. For instance, there are papers about performance that, however, do not target any specific quality features (e.g., safety, sustainability, accessibility) among those emerged from the keywording phase. Lastly, those papers with multiple application fields have been counted multiple times, one per application field. In the following, we briefly report a glossary describing the classes in the first and second level of the taxonomy (i.e., the blue and green classes in Fig. 5), while the remaining ones are self-explanatory.

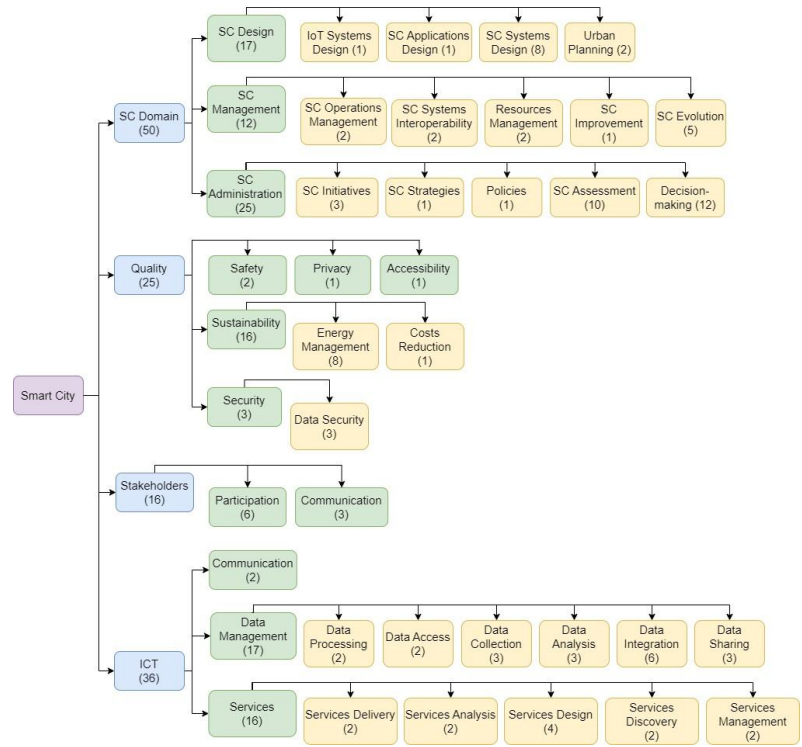

Fig. 5.  Application fields taxonomy.

- **SC Domain** refers to those publications whose application field deals with:
  - *SC Design*, the design of SCs (sub-)systems. Their aim is that of improving the design, planning, and performance of future SCs.
  - *SC Management*, the management of all the subsystems composing a SC to better handle their complexity, thus supporting subsystems' analysis, communication among them, and SC improvement and evolution.
  - *SC Administration*, the administration of SC projects and initiatives aiming to support countries' development, to positively impact on the implementation and success of a SC.
- **Quality**, refers to those publications whose application field regards the supporting of quality features (i.e., safety, privacy, sustainability, accessibility, security) that are relevant in the SC domain.
- **Stakeholders**, refers to those publications whose application field deals with the creation of meaningful social and cultural implications opening new ways for citizens' and stakeholders':
  - *Participation*, to increase interoperability and the active / proactive participation in the SC life.
  - *Communication*, to increase openness, social inclusion, integration, education.
- **ICT**, refers to those publications whose application field deals with:
  - *Communication*, the support of the communication between heterogeneous systems in a SC through the effective management of the different kinds of communication protocols.
  - *Data Management*, the management of all the processes (e.g., data collection, data analysis, data storing) applicable to the data coming from a SC.
  - *Services*, the management and implementation of smart services in SCs.

The taxonomy emerged from our analysis highlights that the majority of the publications belong to the *SC Domain* (**50**), mainly dealing with *SC Administration* (**25**), where almost half of them focus on *Decision-making* (**12**). This is clearly in line with the distribution of publications among the SC dimensions, reported in Fig. 3, where the *Smart Governance* is the predominant dimension. In the *ICT* field (**36**) the *Data Management* (**17**) predominates followed by *Services* (**16**). This is comprehensible given that, SCs and their constituent systems act more and more as (big) data producers, and the offer of value-added services is also a characterising feature of SCs. The *Quality* application field (**25**) follows. In line with the SC concept and its objectives, *Sustainability* (**16**) is the most investigated quality feature. This might reflect the fact that the *Smart Environment* dimension is the second most investigated one, according to Fig. 3. Surprisingly, besides the objective of SCs of supporting social inclusion, integration, education and active citizens' involvement, the *Stakeholders* field (**16**) is only marginally considered when dealing with SCs modeling.

### 4.3 RQ3: Which modeling approaches have been used to represent Smart Cities?

To answer RQ3 we extracted the modeling approaches and techniques used in every publication to model the SC domain or subdomains. We report in Figure 6 the most used model approaches with at least 3 occurrences through a bar chart. We highlight here that those publications using multiple approaches and/or techniques have been counted multiple times in the given bar chart. We can notice that *business*, *architectural*, and *ontology* modeling are the most prominent detected approaches, with **15**, **14** and **12** occurrences, respectively. In particular, *business* modeling approaches essentially include modeling canvas. For instance, in [S41, S50] Business Model Canvas (BMC) are used for the monitoring and development of economic value creation in a smart city. Under *architecture* modeling approaches we mainly grouped reference architectures (e.g., [S2] presenting a reference architecture for SC projects, formalizing the process of creating SC applications and services), IoT-based architectures (e.g., [S93] presenting an Edge-to-Cloud-as-a-service model for building efficient services in a SC domain), big data architectures (e.g., [S44] proposing an architecture to support the management and analysis of huge amount of distributed big data). The third most prominent modeling approach is *ontology*. Ontology-based approaches have been distinguished from *conceptual* modeling based on the explicit use of ontology languages and/or tools (e.g., [S3, S94]). In other words, those publications that make use of conceptual modeling without exploiting ontology technologies have been classified simply as conceptual. *Data* modeling approaches follow, by including, for instance, class diagrams [S4], data meta-model [S52], Linked Open Data (LOD) [S67], data transmission model [S62]. *3D* and *behavioral* modeling approaches share the fifth position. 3D modeling has been used, for instance, in smart building design (e.g., [S23]), energy management systems (e.g., [S36]), training systems based on augmented reality (e.g., [S103]). *Behavioral* modeling approaches include UML models (e.g., state machine [S22, S106], sequence diagram [S72]), Markov chain model (e.g., [S53]) and flowchart [S106]. *Mathematical* modeling approaches deserve a discussion. Despite we understand that they may appear to be outside the scope of this study, we decided to keep them, since they have been used to mainly specify cost models (e.g., [S88, S30]) in different SC context, or for instance, energy management (e.g., [S81]) and waste management (e.g., [S104]) in the sustainability field. Through the end of the classification we can see *evaluation* modeling approaches including essentially KPIs-based models (e.g., [S105, S18]. *Service-based* modeling include approaches for the design, discovery and management of services, in different SC context (e.g., [S27, S19, S63]). *Conceptual* modeling approaches differ from ontology approaches, as stated before. They include, for instance, graphical domain-specific languages (e.g., [S10, S70]), used for representing SC scenarios in a visual still abstract and conceptual way. Under *multi-agent* modeling approaches we classified those approaches explicitly based on the concept of agents (e.g., [S31, S35]). The *maturity* modeling approaches close the classification (e.g., [S21], [S42]).

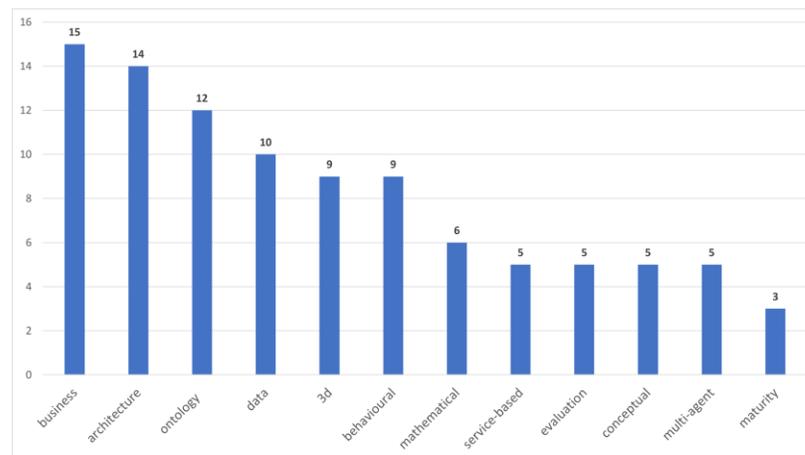

Fig. 6. Number of publications for the most used modeling approaches.

To better understand in which SCs fields these prominent model kinds are used, we further correlated the data about the most used model kinds with the taxonomy of application fields we defined and discussed in Section 4.2.

In Fig. 7 we report the most prominent application fields for the most used modeling approaches in a bubble chart. In the first column of the chart the most prominent application fields for *business* models are reported. Here we can observe a clear predominance of the SC Domain class, followed by the ICT class. Considering that business models represent a mean for organizations to create, deliver, and capture value, in contexts such as economic, social, cultural, this fits with the SC Domain's subclasses, namely SC Management and SC Administration, where business models can indeed be exploited. Furthermore, business models can provide support in the Data Management and Services subclasses of the ICT class, which also emerges among the application fields for this model kind. The second column shows that *architectural* models have been mainly used in the SC Domain, ICT, and Quality classes of our taxonomy. Considering that as architectural models we mainly identified reference, IoT, and big data architectures, this is in line with the prevalence of the SC Domain, which also include SC Design and the ICT class, which further contains the Data Management subclass. The Quality class is also understandable given that architectural models are also used for performance analysis. The third column highlights a prevalence of the ICT and SC Domain application fields for *ontology* models. Specifically, in the ICT field there is a prevalence of the Data Management subclass, while in the SC Domain field, the SC Administration subclass is predominant and the decision-making sub-subclass recurs. These results are perfectly in line with the wide use of ontology for knowledge representation and reasoning, integration and exchange of data and knowledge, and conceptual data modeling.

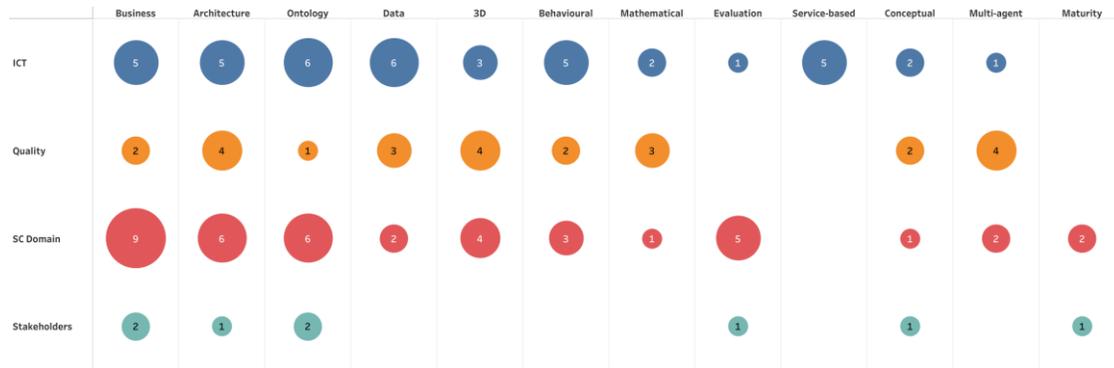

Fig. 7. Correlation analysis between the model kinds and the application fields taxonomy.

The one just described is the correlation analysis for the three most prominent model kinds. By observing the bubble graph, we can notice that the trend emerged for the business, architectural, and ontology modeling approaches is recurring also for the remaining modeling approaches, by confirming the predominance of works belonging to the SC Domain and ICT application fields, followed by Quality and Stakeholders.

## 4.4 RQ4: What is the maturity status of smart cities modeling approaches?

To answer RQ4, we performed three different analysis. Firstly, we classified every publication w.r.t. the research type facets (Section 4.4.1). Secondly, we calculated the technology readiness level of the surveyed works (Section 4.4.2). Thirdly, we made a tool support analysis involving those publications contributing with a tool (Section 4.4.3).

*4.4.1 Research type facets analysis.* As first analysis, we classified every publication w.r.t. the research type facets. As stated in Section 3.4, the research type facets (listed in Table 2) are used to distinguish between empirical and non-empirical papers and to categorize them based on the proposed solution and level of validation. In Figure 8, we can see that the majority of the publications, i.e., around 73%, belongs to the *Solution*, *Validation*, and *Evaluation* types. The *Philosophical* facet follows, with around 24% of publications. As regard the *Experience* and *Opinion* facets, very few publications, i.e., around 4%, fall in this categories. From the results reported in Figure 8 we can infer that half of the publications, namely around the 54%, present a contribution, mostly in the form of a solution to a given problem (i.e., Solution facet), or a taxonomy or a conceptual model (i.e., Philosophical facet). Approximately, only 24% of contributions are validated through experiments in laboratory environment (i.e., Validation facet) and only 19% of them performed an evaluation research, namely through the solution implementation and the implementation evaluation (i.e., Evaluation facet). We further made a trend analysis for the different type facets over the years. Differently than our expectations, we did not observe any particular trend.

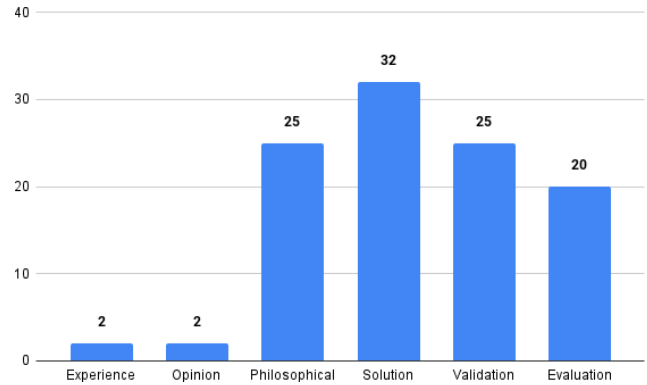

Fig. 8. Number of publications w.r.t. the type facets.

*4.4.2 Technology readiness level analysis.* As a second analysis, we calculated the TRL (Technology Readiness Level) only for those publications classified as *Solution*, *Validation* and *Evaluation*, namely 77 in total. We used the Excel spreadsheet reported in the replication package[8] and described in Section 3.4. The tool calculates the TRL of a contribution w.r.t. the maturity level of the used *Technology*, *Development* and *Design*. Specifically, the maturity level is measured according to a scale ranging from 0 to 5. The overall TRL, instead, is derived as a weighted sum of the *Technology*, *Development* and *Design* maturity levels, and it ranges from 0 to 9, as reported in Table 3.

In Figure 9, we can see that the contributions of type *Solution* have low values for all the three aspects taken into account for the calculation of their TRL. In particular, for the technological and design maturity, besides few outliers, the median values is equals to 2. As regard the development maturity level the median value is equal to 1. These results highlight that great part of contributions of type Solution are in a preliminary phase in which technological and development aspects have been defined but the implementation is still in a preliminary stage.

For the Validation research type facet, we can see in Figure 10 that the maturity level w.r.t. the three considered aspects is higher if compared with those of the Solution papers. In particular, the technological maturity is much higher, with values that go from 3 to 5. Meanwhile, the development and design maturity levels are in the middle, ranging from 1 to 4 with both the medians equals to 2. In this case, the contributions are technologically more mature, since they have already begun to be tested in a laboratory environment, although their maturity is lower w.r.t. the design and development, since they are not in their final form yet. The Evaluation facet in Figure 11 presents slight increases in the values for the three maturity levels, compared to the Validation facet. In particular, the technological maturity has values going from 3 to 5. For the development maturity level we have values going from 2 to 4 with the median equal to 4, thus quite larger compared to the validation facet. The design maturity is, instead, rather constant around the value of 3. Even if the contributions in the Evaluation facet have a higher technological level, meaning that the works are in their near final form, great part of them have not yet been distributed in the real world but only tested in a sub-set of conditions and environments. Eventually, the distribution of the calculated TRL values for the three facets are reported in Figure 12. We can observe that for the solution facet we have mostly TRL values that are around 2, with the exception of a very few works showing a TRL equals to 1 and 3. For the validation facet, the TRL goes mostly from 3 to 8, with the median equals to 5. Lastly, the evaluation facet has TRL values mostly of 7 and 8. Interestingly, we can notice here that no contribution shows a TRL value of 9, namely presenting an actual system proven in operational environment, meaning that we have not found mature and stable systems or framework.

---

[8]  https://drive.google.com/file/d/15uutEqTJeyqjmJRL_OxkbFrstYdgVnjG/view?usp=share_link

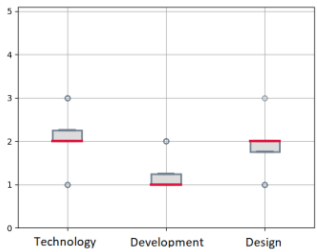

Fig. 9. Boxplot with the maturity level for the facet Solution.

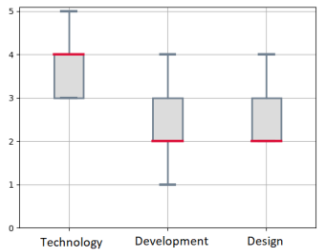

Fig. 10. Boxplot with the maturity level for the facet Validation.

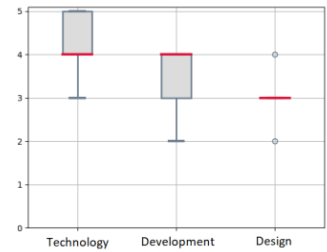

Fig. 11. Boxplot with the maturity level for the facet Evaluation.

*4.4.3 Tool support analysis.* In order to reply to RQ4, we performed a third analysis to check those contributions presenting *proprietary tools* or *open source solutions*. Overall, we found only 15 publications over 106 with an explicit reference to a developed tool. In Figure 13 (left side) we can see that 9 publications refer to an open source solution whereas 6 publications to a proprietary tool. Moreover, in Figure 13 (right side) we report the distribution of the publications with the evidence of a tool over the research type facets. As expected, they belong only to the Solution, Validation and Evaluation facets, those collecting the most mature works. We can further observe that, reasonably, most of them fall into the Evaluation facet, which brings together the most technologically mature solutions.

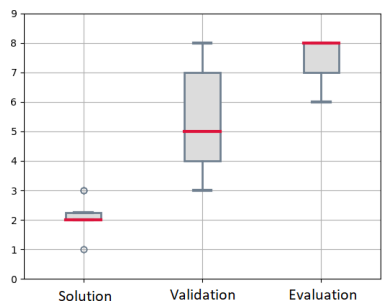

Fig. 12. Boxplot with the TRL values w.r.t. the three facets.

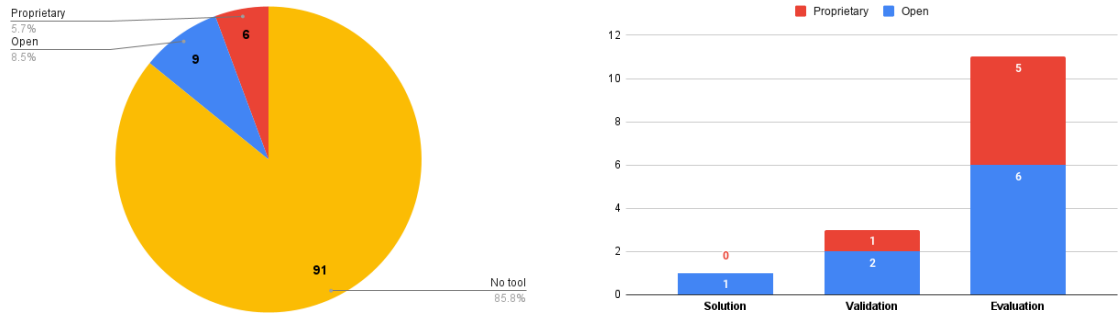

Fig. 13. (**LeP side**) Overview of publications contributing with open or proprietary tools. (**Right side**) Distribution of publications contributing with open or proprietary tools over the Solution, Validation and Evaluation facets.

As regards the 9 papers contributing with an open tool solution, we further analyzed them and report details in Table 4. In particular, for every publication we specify the following attributes:

- *Paper ID*, the reference of the paper as listed in the corresponding replication package[9];
- *Dimension*, the SC dimension, i.e., smart environment (SE), smart living (SL), smart governance (SG), smart people (SP), smart mobility (SM), smart economy (SEc), and smart things (ST);
- *Tool Type*, namely the type of the contribution (e.g., Web platform, Android application, etc.);
- *Details*, i.e., additional information about the proposed tools;
- *Category*, specifying if the tool is developed for a specific scenario (*Specific*) or it is applicable in more environments (*General*);

| Paper ID | Dimension | Tool Type | Details | Category |
|----------|-----------|-----------|---------|----------|
| [S6] | SE | Web Application | Air pollution monitoring. | General |
| [S7] | SL | Standalone Platform | Energy load profiles creator. | General |
| [S17] | SG | Web Application | Wellbeing benchmarking of Canadian communities. | Specific |
| [S42] | SG | Web Application | SCs maturity assessment and benchmarking. | General |
| [S49] | SM | Mobile Application | Drivers Monitoring (e.g., driving parameters and biometrics information). | General |
| [S52] | SG | Standalone Platform, Mobile Application | Interdisciplinary instant point observations. | General |
| [S67] | SG | Web Application | Web portal for the municipality of Catania (Italy). | Specific |
| [S96] | SL | Web Application | Development and storage of 3D virtual city model. | Specific |
| [S100] | SL | Standalone Platform | Energy Management Benchmarking. | General |

Table 4. Details of publications contributing with an open tool.

Looking at the results in Table 4, we can observe that the open tool solutions span over 4 of the 7 SC dimensions, namely Smart Environment, Smart Living, Smart Governance, and Smart Mobility, with a distribution showing a peak for the smart governance dimension (i.e., 4 tools over 9), in line with the results reported in Figure 3, about the most modeled SC dimensions. The majority of the tools are presented as Web applications (i.e., Web platforms, Web frameworks, Websites), while the remaining ones span from Python platform to Android application. Moreover, they are typically open for applicability in different environments (i.e., they do not rely only on one or a few SCs; to the contrary, they are indistinctly applicable to different SCs or in diverse geographical areas), namely *General*. In particular, these kinds of tools represent useful instruments for the cities, referring to aspects which are not strictly connected with the city itself (e.g., air pollution monitoring tools, SCs assessment and benchmarking tools), and they can be easily used in different SCs. However, some of them are classified as *Specific*; this happens mainly when the tools refers to a particular SC (or cities of a given geographical region), as can be clearly observed in column Details of Table 4. Differently than general tools, Specific tools are highly coupled with the city they refer to (e.g., the city structure or infrastructures), thus limiting the tool applicability. To conclude the discussion on **RQ4**, we must highlight here that we have been not able to use and analyse those proprietary tools for which we do not have access. All the findings we spotted about them come only from the corresponding publications. We are aware that this can limit our analysis, by hindering the evaluation of contributions. However, given the limited number of publications contributing with a proprietary tool, we can state that whatever their level of maturity, this would not alter our mapping results significantly.

### 4.5 RQ5: When did the contributions on modeling Smart Cities occur?

To answer RQ5, we analysed the distribution of publications over the years. In Figure 14, we can see that the first contributions on the topic appeared in 2011. Reasonably, it might be that the research about modeling SCs started simultaneously with the advent of the concept pointing out the smartness of cities [31]. However, a significant increase of publications started from 2015 and it remained quite constant up today, apart some fluctuations.

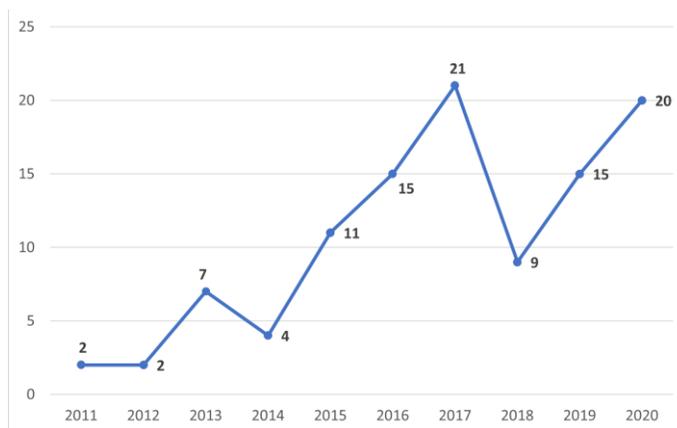

Fig. 14. Number of publications per year.

Moreover, we made a further analysis by differentiating between *Article*s in scientific journals and *InProceedings* publications in scientific conferences and workshops. In Figure 15, we report the number of publications per year w.r.t. the publications type. We can notice that inProceedings publications appeared in 2013, and from 2014 the they started to be increasingly much more than articles. Noteworthy, in 2018 we can observe a significant decrease for both publications types, while in 2020 a change in the trends of the two series can be seen, i.e., inProceedings publications started decreasing while articles started increasing. This might depend on the fact that contributions on the topic are achieving higher quality and maturity over the years, leading to valuable publications. Anyhow this is just an assumption, and future observations are required.

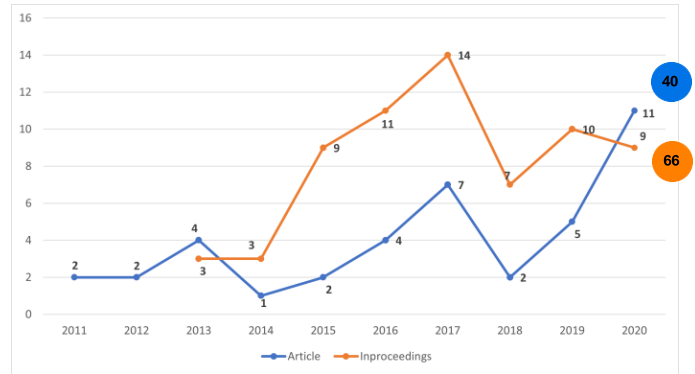

Fig. 15. Number of publications per year w.r.t. the publications type.

### 4.6 RQ6: Where have the contributions been published?

To answer RQ6 we analysed the venues of both types of publications, i.e., inProceedings and articles. We already know from Figure 15 that the number of inProceedings publications, namely 66, is higher than the articles one, namely 40. Starting from analysing inProceedings publications, we found more than 53 different conference venues. We report in Table 5 only the most prominent conferences, i.e., those with at least three publications. In Table 5 we indicate, for each conference or workshop, the *Venue* name with its acronym, the information about the corresponding *Scientific Area*[10], and the number of *Publications* to that venue. We identified 4 prominent conference venues and we discuss them in the following. The main venue is the *IEEE International Smart Cities Conference (ISC2)* with 4 publications. This conference relates to the *Computer Science* and *Social Sciences* scientific areas and this result is totally in line with the interdisciplinary aspect of the SC domain. The second most prominent venues are 3, each with 3 publications. The *Smart Cities Symposium (SCSP)* that has a multidisciplinary perspective, at the crossroads of *Computer Science*, *Energy* and *Decision Sciences*. The *International Conference on Smart Cities and Green ICT Systems (SMARTGREENS)*. This conference is focused more on the ICT aspects in the SC domain, this is why it is more *Computer Science* oriented. The *Winter Simulation Conference (WSC)* in the scientific areas of *Computer Science, Engineering*, and *Mathematics*. The remaining venues show also multidisciplinary scientific areas, and report a number of publications less or equal than 2.

For publications of type articles, we found more than 32 different journals venues. In Table 6 we report the most prominent journals, i.e., those with at least 2 publications. For journals venues, besides the attributes used for conferences, namely *Venue*, *Scientific Area*, and *Publications*, we further indicate the *Quartile* assigned to each journal, i.e., the journal impact index measuring the quality of scientific journals, ranging from Q1 (the highest ranked journals) to Q4 (the lowest ranked journals). We remind here that we kept only journals ranked Q1 or Q2, according to our inclusion criteria.

---

[10] We retrieved the scientific area from *Scimago* https://www.scimagojr.com/, i.e., a publicly available portal that provides journals and conferences rankings. Last access on 11 October 2023.

| Venue | Scientific Area | Publications |
|---|---|---|
| ISC2 (IEEE International Smart Cities Conference) | Computer Science; Social Sciences | 4 |
| SCSP (Smart Cities Symposium) | Computer Science; Energy; Social Sciences | 3 |
| SMARTGREENS (International Conference on Smart Cities and Green ICT Systems) | Computer Science | 3 |
| WSC (Winter Simulation Conference) | Computer Science; Engineering; Mathematics | 3 |

Table 5. Most prominent conferences regarding the number of publications with at least 3 publications.

We identified 7 prominent journals and we discuss the first two in the following. The main venues are *ISPRS International Journal of Geo-Information* and *IEEE Access* with 3 publications each, fitting into the *Q1* quartile. The first belongs to the *Earth and Planetary* and *Social Sciences* scientific areas. Instead, the second one to *Computer Science*, *Engineering*, and *Material Sciences*. Looking at the *Scientific Area* column, also journals refer to interdisciplinary communities.

| Venue | Scientific Area | Quartile | Publications |
|---|---|---|---|
| ISPRS International Journal of Geo-Information | Earth and Planetary Sciences; Social Sciences | Q1 | 3 |
| IEEE Access | Computer Science; Material Sciences; Engineering | Q1 | 3 |
| Journal of Urban Technology | Social Sciences | Q1 | 2 |
| IEEE Communications Magazine | Computer Science; Engineering | Q1 | 2 |
| Energies | Energy; Engineering; Mathematics | Q2 | 2 |
| Wireless Personal Communications | Computer Science; Engineering | Q2 | 2 |
| International Journal of Electronic Government Research | Computer Science; Social Sciences | Q2 | 2 |

Table 6. Most prominent journals regarding the number of publications with at least 2 publications.

Lastly, we can observe that both journals and conferences span across multiple scientific areas and interdisciplinary venues. Furthermore, *Computer Science* is the prominent scientific area.

## 5 MAIN FINDINGS AND RESEARCH DIRECTIONS FOR MODEL-DRIVEN ENGINEERING

In this paper, we performed a systematic mapping study to identify the relevant research about modeling approaches for SCs. In Section 3.1, we grouped the relevant research questions w.r.t. problem, solution, and scientific community domains. In the following, we give the summary of our main findings and we present the envisaged research directions we have identified. To do so, we maintain the same distinction among the problem, solution, and scientific community.

### 5.1 Problem Domain

In the problem domain, we identified the research questions **RQ1** and **RQ2** that are about the most prominent modeled SC dimensions and application fields in the SC domain, respectively. As discussed in Section 4.1, SCs can be defined in terms of multiple dimensions. In particular, we used the Boyd Cohen Wheel standard [12]. We found out that the *smart governance*, *smart environment* and *smart people* have mostly elicited the research interest, from the modeling perspective. With respect to the application fields of SCs modeling approaches, we derived a corresponding taxonomy (Section 4.2). It shows that SC modeling has been largely exploited in the *SC Domain* field, which includes *SC Design*, *SC Management* and *SC Administration*. *ICT* and *Quality* fields follow, while *Stakeholders* closes the ranking. Based on our main findings in the problem domain, we outline the following research directions.

**Smart cities dimensions: uneven coverage.** Although SCs are organized in different dimensions, which support also a better comprehension and management, such dimensions are not meant to be tightly separated. Indeed, each of them may relate to one or more other dimensions. For instance, realizing smart mobility might support smart environment. With these premises, we believe that when referring to SCs modeling, a full coverage of all the dimensions is required. Of course, given the complexity of the domain, the different dimensions can be treated separately by following a divide and conquer approach. However, with the ambition to converge to a reference model for SCs, we find it as important that research progresses in each SCs dimension. In this study, by only considering the number of papers per dimension, we can spot that, differently from the governance and environment dimensions, the coverage for all the remaining ones seems to be currently limited when it comes to modeling support. For instance, regarding the *smart economy*, we believe that it requires specific competencies and background knowledge. In addition, economy is a area where mathematical modeling is of major relevance. However, the limited coverage of this dimension in this study might be due to a corresponding lack of dedicated modeling languages for smart cities. Lastly, w.r.t. *smart things*, the limited coverage we observed in our results may be due to the fact that modeling approaches for this dimension have been published detached from the SC domain and with a stronger focus on IoT. To conclude, by considering the interrelation among SCs dimensions and the results of this study, we argue that as a first step researchers should make a reflection about if and to which extent different modeling approaches are needed for the different dimensions. Still, an open question is how to integrate the different dimensions, e.g., by developing a family of languages for SCs modeling.

As future research directions, we believe it is of relevant importance to deeply investigate the reasons of this uneven coverage of the SC's dimensions. To this aim, it is a prerequisite to better understand what each dimension is about and how much it is overlapping with the others. This means that we first need to reach a global overview of the SCs domain and dimensions in order to be able to establish *best practices* for modeling them. Specifically, this would help to identify: (*i*) which are the requirements the models should satisfy? (*ii*) what do we have to model, e.g., level of detail? (*iii*) what is the purpose of the model?, and, (*iv*) to whom the model is addressed? Finally, considering also the multi-disciplinary nature of SCs, we believe that it is crucial to promote inter- and intra-communities interaction to overcome the current limited coverage of some SC dimensions compared to others as well as to identify the interfaces between the different dimensions.

**Smart cities application fields: more stakeholders inclusion need.** As discussed in this paper, one of the goals of SCs is to promote their sustainable development to improve human lives and preserve the environment. To this aim, multiple dimensions (e.g., people, mobility, economy) come into play, together with their corresponding stakeholders (e.g., citizens, private companies, public administrations). With these premises, our taxonomy about the application fields of SC modeling approaches highlighted that the *Stakeholders* field, including *Participation* and *Communication* sub-fields, emerges with 16 works over 106 that have been covered by this study, i.e., a marginal subset. This might be a sign of a lack of approaches aiming directly at supporting the inclusion, participation, and communication with stakeholders. If this observation is confirmed, it would highlight a serious shortcoming, considering the crucial importance of the stakeholders inclusion for effectively realizing SCs. Indeed, stakeholders are diverse and they can play crucial roles in diverse SCs realization phases: from the SCs design, which need the involvement of authorities, governance, public and private companies, etc., to the SCs daily lives, which must include citizens, tourists, operators, etc. With these premises, we argue that the stakeholders inclusion must move from a marginal to a central setting.

> As future research direction, we believe that our findings provide a foundation for a deeper analysis regarding the support offered through model-driven approaches to the stakeholders inclusion in SCs environments. In other words, it is worth to further investigate the occurrences, relevance and diffusion of (model-driven) solutions supporting inclusion, participation, communication and knowledge sharing devoted to SC stakeholders, to effectively claim the state of significance of the provided stakeholders support. More context-sensitive research may be required to learn more about the potential users and their backgrounds for modeling approaches in the SC domain. In this context, Low-Code Development Platforms (LCDPs) [35, 36] may play a relevant role. LCDPs are visual environments, commonly available on the Cloud or as Web-based applications (which seems to be a good fit for SCs), that are being increasingly introduced and promoted for citizen developers. Specifically, in the context of SCs modeling, they would enable stakeholders without any particular programming background to take an active part in the modeling of the SCs they refer to, given their role and knowledge with which they can contribute.

## 5.2 Solution Domain

In the solution domain, we identified the research questions **RQ3** and **RQ4** that are about the modeling approaches currently used to design SCs and SCs subsystems, and their status in terms of maturity, respectively. As discussed in Section 4.3, *business*, *architectural*, and *ontology* modeling are the most prominent approaches, and they are especially used in the *SC Domain* and *ICT* application fields. Regarding the research status in SC modeling, we performed multiple analysis (Section 4.3). We classified publications according to the research type facets, to distinguish empirical and non-empirical papers and to categorize them based on the proposed solution and level of validation. This analysis turned out that the majority of the publications, i.e., around 73%, belongs to the *Solution*, *Validation* and *Evaluation* types. Specifically, only 24% of the contributions are validated through experiments in laboratory environment and only 19% performed an evaluation research through solution implementation and evaluation. In terms of maturity, we calculated the TRL for those publications classified as *Solution*, *Validation* and *Evaluation*, which measures the TRL of a contribution w.r.t. the maturity level of the used technology, development, and design. In summary, no contribution shows a TRL value of 9, namely presenting an actual system proven in operational environment, meaning that this validation has still to be performed in SC modeling. Moreover, only 15 publications over 106 contribute a developed tool (open-source or proprietary). Based on our findings in the solution domain, we derive the following research directions.

**Modeling smart cities: wide variety of modeling paradigms.** From our findings, it can be concluded that a uniform way for modeling SCs does not yet exist. Indeed, the reference modeling approach or framework has not yet emerged. This might be due to the fact that researchers and practitioners tend to model only those aspects relevant for their specific purpose, by exploiting the most fitting modeling paradigm, such as business, architectural, ontology, and so on. As a matter of fact, given the complexity and multi-disciplinary nature of the SCs domain, it is reasonable to focus on specific aspects, dimensions, subsystems, and, then, on the most appropriate modeling approaches following the main idea of Multi-Paradigm Modeling [5, 43] based on the research goals one is pursuing.

> As future research direction, we believe that it might be useful for the involved research communities to define a hybrid solution in SCs modeling. The idea is to overcome the divergence arising from the multitude of models used to design SCs and their different dimensions and subsystems. To this aim, after a decade of research in SCs modeling, further investigations should be undertaken for abstracting from the knowledge we gain as a community on the topic and derive a hybrid solution to harmonise the knowledge and models, with the aim of converging to a reference solution, which may become a standard in the long term. This way, such solution might be used for customization to model different SC aspects for different needs. Alternatively, it might be valuable, by learning from the emerged modeling approaches or relevant solutions identified also in this study, to provide a multi-viewpoint approach allowing to keep the different formalisms for specific purposes, while aiming to identify and maintain the relationships between the different views or transformations from one view to the other.

**Maturity of contributions: MDE to the rescue.** From our analysis, it can be derived that only few works contribute with technologically mature solutions. Specifically, we observed that more than half of publications, namely around the 54%, present a contribution mostly in the form of a solution to a given problem, not yet evaluated (i.e., Solution facet), or a taxonomy or a conceptual model (i.e., Philosophical facet). In addition, the TRL measurement has shown that for the presented modeling approaches the evaluation in operational environments is currently missing. This is further highlighted by the low number of papers (i.e., 15 out of 106) contributing a dedicated tool. Of course, we are aware that other means may be more suitable to search for solutions with a TRL of value 9, e.g., a multi-vocal review would be promising to deeper investigate the maturity status of existing solutions. Nevertheless, our results show that more than half of the surveyed publications stay solely at a conceptual level, without showing any validation or evaluation.

> As future research directions, we believe that the transition from research prototypes towards industry-scale solutions is an important short/mid term objective for providing useful instruments for SCs. To this aim, MDE can come into play with its body of existing technologies that have been already proven useful in other industrial settings, e.g., see Industry 4.0 [46]. While this may be a long-term goal, considering also our discussion on the suitability of searching for solutions with TRL 9 in research works, an alternative might be the following. According to the technology maturation model introduced by Mary Shaw [38], the path through more solid and stable solution can be gradual, by progressing through different phases, namely from research-oriented phases to development and commercialization phases requiring the involvement of industries to reach the highest maturity level. This alternative looks reasonable and the industrial support might clearly lead to solutions with higher TRL values w.r.t. those observed in research works. With these premises, we can argue that there are the conditions under which MDE can actually take place and give the boost that research needs in the SCs modeling context. Indeed, MDE technologies may evidently support the development of modeling languages and their environments, as well as their management, such as comparison, model transformation, code generation.

### 5.3 Scientific Community Domain

In the scientific community domain, we identified the research questions **RQ5** and **RQ6** that are about when and where the contributions on modeling SCs occur, respectively. In Section 4.5, we discussed that the research about modeling SCs started in 2011. In 2020, a change in the trends of the two series of contributions, i.e., *inProceedings* and *article*, can be seen, i.e., inProceedings publications started to decrease while articles started to increase. This might indicate that the contributions on the topic are maturing over the years. In Section 4.6, we analysed the venues of both types of publications. The inProceedings contributions span over 53 different conference venues, among which the most prominent is the *IEEE International Smart Cities Conference (ISC2)*, with four publications on modeling SCs. However, in addition to ISC2, there are other venues targeting specifically the SCs domain. This might suggest that a general community on smart cities exist which is interested in modeling aspects. However, at the same time, this topic is also spreading among further communities and venues beyond core SC events, e.g., see the Winter Simulation Conference (WSC) which highlights the importance of modeling and simulation in the SC domain. The articles contributions span over 32 different journals venues without a clear main target journal for SC modeling research. Lastly, both paper types refer mostly to interdisciplinary communities, in line with the multi-disciplinary nature of SCs. Based on our findings in the scientific community domain, we derive the following research directions.

**Modeling as a cross-cutting community concern.** With our analysis, we found out that there is a wide distribution of different conferences and journal venues where contributions on modeling SCs have been published. On the one hand, this might be due to the fact that the research on this topic is conducted by researchers from interdisciplinary communities, which have their dedicated publication venues. On the other hand, by also considering the contributions in the different scientific areas, separately (e.g., the Computer Science area), it does not emerge a reference community for SC modeling. However, as discussed above, and to emphasize that modeling is a cross-cutting concern, we highlight that there are specific venues for SCs (e.g., ISC2, SCS, ICSC), as well as other conferences (e.g., WSC, ER) focusing on general techniques, such as modeling and simulation, which are covering contributions pertaining to the SCs domain. On the one hand, this proliferation of SC modeling approaches spreading over many different venues might be seen as counter-productive, by further hindering the communication among the different researchers coming from the different research communities. On the other hand, it might be seen as evidence that there exist dedicated communities efficiently discussing specific SC modeling topics. However, there are also cross-cutting concerns, such as technologies, languages, frameworks, that would require more interactions among these communities, in the best case, discussions between domain experts as well as technology experts.

**The role of the Model-Driven Engineering community.** The overall objective of the study is to identify research trends and analyze the characteristics of SC modeling contributions, by means of several classifications. The results and main findings of this study are expected to serve as useful guidelines for researchers and practitioners working on SC projects. In particular, given the nature of the investigated approaches, specifically based on modeling techniques, we believe that our findings can be particularly relevant and influential to the Model-Driven Engineering (MDE) community. According to our findings highlighting several gaps and lacks in terms of modeling smart cities, it clearly emerges that the adoption of the body of knowledge of the MDE community [10] may be further improved.

Finally, we recommend general directions, by combining the single findings discussed above. More precisely, we believe that the MDE community is an enabling community for SCs to converge research efforts to provide a *reference model* for SCs. More precisely, given the multi-disciplinary nature of SCs, also confirmed by this study, such reference model should offer a *multi-viewpoint modeling* perspective, to accommodate the different views and the diverse stakeholders typically involved in the SC domain. Moreover, *tool supported language definitions and workbench development* by the MDE community might be helpful to further improve the technological maturity level of existing approaches. In this regards, the reference model we envision must be abstract enough to consider *different technical spaces* such as open linked data, multi-agent systems, and web services to mention just a few, given the heterogeneous nature of SCs, as well as being able to enable the involvement of different domain experts by concrete language support.

## 6 THREATS TO VALIDITY

To assess the validity of our study, we discuss here the four basic threats to validity according to the suggestions by Wohlin *et al.* [45], and we argue about how we behaved in order to mitigate them.

*Construct validity.* Construct validity is concerned with the generalization of the experiment result to the theory behind the experiment [45]. It essentially refers to the extent to which the experiment setting reflects the theory. It deals with issues arising during the research design phase and issues related to personal guessing of the experimenters.

In the case of our study, among the *design threats* listed in [45], it might be biased by: (i) the so-called *mono-method bias*, which refers to the fact of using a single type of measures or observations; (ii) the *confounding constructs and levels of constructs*, which refers not primarily to the presence or absence of a construct, but on the level of a construct which is of importance to the outcome. Specifically, the former threat may be due to the fact that we followed the systematic mapping process by Petersen *et al.* [29]. To mitigate it, we formed two independent groups of two authors, each one analysing the work of the other and discussing the obtained results, to avoid observation bias, thus misleading results. The latter, instead, applies mainly to the fact that we used several classifications in order to categorize the papers, e.g., research facets, SC dimensions. However, a paper might belong to more than one facet or dimension. In this case, the mitigation strategy was again supported by our organization in two independent groups. A group of authors performed the classifications, by discussing all those papers for which multiple facets or dimensions could be applied, and selecting the most fitting one. The other group, subsequently, revised the classifications of a consistent subset of the surveyed papers. In case of conflicts or misalignment, a discussion among all the authors was performed.

Eventually, w.r.t. social threats, experimenter expectancies can bias the study outcome. Although all of us had some expectations about the outcome of the study, probably due to our knowledge of the SCs and/or model-driven engineering domains, we mitigated this threat by letting the mapping emerge by means of open and replicable analysis, and by providing new insights, such as the taxonomy in Figure 5.

*Internal validity.* Internal validity concerns with the causal relationship between the treatment and the outcome [45]. In other words, it must be avoided that this relationship is affected by external factors out from the experimenters control. In the case of our mapping study, threats to internal validity may come essentially from the study identification, data extraction and classification, and they may affect the quality of the sample of obtained studies.

As concern the study identification, we defined a clear search string, by considering synonyms (i.e., model vs. modeling vs. modelling) as well as singular and plural forms (i.e., smart city vs. smart cities). Only the title of publications were searched for the search string keywords, to make an accurate search and getting more precise results effectively focused on the use of modeling approaches in the SCs domain. Although we used well-known digital libraries to identify the candidate papers (i.e., ACM, IEEE, Scopus, DBLP), they may have a bias towards Computer Science, while other libraries specifically from other research fields have not been investigated. However, this bias is mitigated by the use of the Scopus library that, as mentioned on its homepage, "provides a comprehensive overview of worldwide research output in the fields of science, technology, medicine, social sciences, and arts and humanities." Before downloading the final datasets, we practiced with the libraries to learn their different queries syntaxes and be sure we were using them properly. The inclusion and exclusion criteria are also an instrument to mitigate threats to the study identification validity. Therefore, inclusion/exclusion criteria, the search string and the digital libraries to exploit, were clearly discussed by the authors. All the authors verified the datasets and the application of the criteria for all the borderline papers, such that to avoid a bias due to analysis performed by a single author. Furthermore, we let the time frame of the survey (i.e., 2011 – 2020) emerge, without applying any constraint on the publication year of the studies.

As regards the data extraction and classification, these may be biased by the experimenters judgement and personal opinion. The mitigation strategy to reduce this threat consisted again in the organization in two independent groups, one performing the extraction and classification by discussing borderline papers, and the second one checking and verifying the results. Intra and inter groups discussions have been made, to reach a final agreement. Of course, a publication bias always remains, since we have no guarantee that all relevant studies were selected. It is possible that some relevant studies were not chosen throughout the search process. Anyhow we believe that the selected search string and the overall systematic approach we followed have mitigated this threat.

*External validity.* External validity refers to the *generalizability* of the study, that is the degree to which we can generalize the results inside or outside of the studied population [45]. To this aim, we remind that we utilized a systematic way to identify the studies and extract the results, through continuous interactions of two independent groups of experimenters discussing and clarifying unclear parts, thus to avoid potential pitfalls. With these premises, we believe our work provides a population sample that can be generalized inside the studied population.

As concern the generalizability of the results outside the studied population, we recall that the study was designed for modeling approaches in the SCs domain. Given the wide scope of the study, we argue that our results might apply to other related populations, with the exclusion of those populations involving modeling approaches out from the model-driven engineering umbrella, as for instance machine learning models, even if applied in the SCs domain.

*Conclusion validity.* According to [45], threats to conclusion validity are concerned with issues that affect the ability to draw the correct conclusion about relations between the treatment and the outcome of an experiment. We must make sure that observations are described accurately and objectively. The systematic mapping process we followed [29] partially supports the accuracy of the study, since it implies clear procedures for the selection, screening, classification and data extraction. The continuous interaction and debate among the authors also aimed to mitigate this issue.

The objectivity of the results, instead, is in part supported by the quantitative data collection and visualization we performed. Moreover, to mitigate this issue, we classified the papers iteratively, in the sense that after the first pass of the keywording step (see Section 3.4), we went through the classification several times in order to adjust and refine it according to the considerations made after a first observation of the emerged results. This allowed us to discuss and revise erroneous classification especially for borderline papers or for papers which do not provide enough information to allow meaningful keywords to be chosen at a first attempt.

## 7 CONCLUSIONS

In this article, we report on our findings regarding the investigated topic on modeling SCs over the last decade by performing a systematic mapping study. We performed our analysis by looking at the problem, solution and scientific community domains. It turned out that modeling approaches have been used for all the SCs dimensions highlighted by the Boyd Cohen standard [12], although with different extents, namely with a prevalence for *Smart Governance* and slightly less for *Smart Things* (**RQ1**). Multiple application fields emerged from the surveyed contributions, from the *SC Domain* to the *Stakeholders* passing through *ICT* and *Quality* (**RQ2**), emphasizing the multi-disciplinary nature of the SCs topic. Many different modeling approaches have been used, especially *business*, *architectural*, and *ontology* models (**RQ3**), although only a minority of works contribute with solid solutions implemented and validated in laboratory or real environments (**RQ4**). Contributions on modeling SCs started in 2011 and, from there, they almost constantly increased over the years (**RQ5**). Lastly, we observed that a reference conference and/or journal venue for this specific topic do not exist (**RQ6**). To the contrary, the 106 surveyed contributions span over 53 different conferences and 32 different journals. This shows that research on modeling approaches for smart cities is currently fragmented and may also benefit from more involvement of the model-driven engineering community.

With these findings, and by taking into account the threats to validity that may affect our analysis, we derived some research directions and suggestions for the interested researchers and practitioners that are worth to be investigated and verified. To this aim, we believe that our findings provide a good foundation for deeper analysis about any aspects related to modeling SCs. In addition, as future work, a multi-vocal review to contrast our findings may be of interest. It would be promising to investigate the maturity status of solutions for SCs modeling used in practice. Indeed, considering that SCs are a practical topic, including grey literature (e.g., blog posts, videos, Web sites and white papers) may support us to discover more further modeling solutions existing in the realm of SCs.

## SURVEYED PAPERS

[S1] J. Zhao, Y. Wang; Toward domain knowledge model for smart city: The core conceptual model; 2015; IEEE First International Smart Cities Conference (ISC2)

[S2] M. Abu-Matar, R. Mizouni; Variability Modeling for Smart City Reference Architectures; 2018; IEEE International Smart Cities Conference (ISC2)

[S3] T. Qamar, N. Z. Bawany, S. Javed, S. Amber; Smart City Services Ontology (SCSO): Semantic Modeling of Smart City Applications; 2019; Seventh International Conference on Digital Information Processing and Communications (ICDIPC)

[S4] G. Anadiotis, E. Hatzoplaki, K. Tsatsakis, T. Tsitsanis; A data model for energy decision support systems for smart cities: The case of BESOS common information model; 2015; International Conference on Smart Cities and Green ICT Systems (SMARTGREENS)

[S5] L. Xia, D. Siyi, W. Shuhua; Research on intellisense information service oriented to value network model in smart city; 2016; IEEE International Conference on Information and Automation (ICIA)

[S6] Q. H. Cao, I. Khan, R. Farahbakhsh, G. Madhusudan, G. M. Lee, N. Crespi; A trust model for data sharing in smart cities; 2016; IEEE International Conference on Communications (ICC)

[S7] I. Stoyanova, E. Gümrükcü, A. Monti; Modular modeling concept and multi-domain simulation for smart cities; 2017; IEEE PES Innovative Smart Grid Technologies Conference Europe (ISGT-Europe)

[S8] L. Abberley, N. Gould, K. Crockett, J. Cheng; Modelling road congestion using ontologies for big data analytics in smart cities; 2017; International Smart Cities Conference (ISC2)

[S9] P. Moos, M. Svitek, Z. Votruba; Smart cities, multi-system approach to system modelling; 2016; Smart Cities Symposium Prague (SCSP)

[S10] D. Bork, R. Buchmann, I. Hawryszkiewycz, D. Karagiannis, N. Tantouris, M. Walch; Using Conceptual Modeling to Support Innovation Challenges in Smart Cities; 2016; IEEE 18th International Conference on High Performance Computing and Communications, IEEE 14th International Conference on Smart City, IEEE 2nd International Conference on Data Science and Systems (HPCC/SmartCity/DSS)

[S11] O. Pribyl, M. Lom, P. Pribyl; Smart charles square: Modeling interconnections of basic building blocks in smart cities; 2017; Smart City Symposium Prague (SCSP)

[S12] C. Cabrera, G. White, A. Palade, S. Clarke; The Right Service at the Right Place: A Service Model for Smart Cities; 2018; IEEE International Conference on Pervasive Computing and Communications (PerCom)

[S13] K. Hashimoto, K. Yamada, K. Tabata, M. Oda, T. Suganuma, A. Rahim, P. Vlacheas, V. Stavroulaki, D. Kelaidonis, A. Georgakopoulos; iKaaS Data Modeling: A Data Model for Community Services and Environment Monitoring in Smart City; 2015; IEEE International Conference on Autonomic Computing

[S14] I. Torre, I. Celik; A model for adaptive accessibility of everyday objects in smart cities; 2016; IEEE 27th Annual International Symposium on Personal, Indoor, and Mobile Radio Communications (PIMRC)

[S15] N. Tcholtchev, P. Lämmel, R. Scholz, W. Konitzer, I. Schieferdecker; Enabling the Structuring, Enhancement and Creation of Urban ICT through the Extension of a Standardized Smart City Reference Model; 2018; IEEE/ACM International Conference on Utility and Cloud Computing Companion (UCC Companion)

[S16] A. Taherkordi, F. Eliassen; Scalable modeling of cloud-based IoT services for smart cities; 2016; IEEE International Conference on Pervasive Computing and Communication Workshops (PerCom Workshops)

[S17] D. Cowan, P. Alencar, K. Young, B. Smale, R. Erb, F. McGarry; A model for the socially smart city practical uses of city-level socio-economic indicators; 2017; IEEE International Conference on Big Data (Big Data)

[S18] Y. Yuan, T. Liu; Evaluation Model and Indicator System of Informationization Applications and Services in Smart Cities; 2014; International Conference on Intelligent Environments

[S19] C. Cabrera, S. Clarke; A Self-Adaptive Service Discovery Model for Smart Cities; 2019; IEEE Transactions on Services Computing

[S20] X. Feng, E. S. Dawam, S. Amin; A New Digital Forensics Model of Smart City Automated Vehicles; 2017; IEEE International Conference on Internet of Things (iThings) and IEEE Green Computing and Communications (GreenCom) and IEEE Cyber, Physical and Social Computing (CPSCom) and IEEE Smart Data (SmartData)

[S21] M. A. Juniawan, P. Sandhyaduhita, B. Purwandari, S. B. Yudhoatmojo, M. A. A. Dewi; Smart government assessment using Scottish Smart City Maturity Model: A case study of Depok city; 2017; International Conference on Advanced Computer Science and Information Systems (ICACSIS)

[S22] M. Nakamura, L. D. Bousquet; Constructing Execution and Life-Cycle Models for Smart City Services with Self-Aware IoT; 2015; IEEE International Conference on Autonomic Computing

[S23] K. Sugihara, Z. Shen; Automatic generation of 3D house models with solar photovoltaic generation for smart city; 2016; 3rd MEC International Conference on Big Data and Smart City (ICBDSC)

[S24] P. Smiari, S. Bibi; A Smart City Application Modeling Framework: A Case Study on Re-engineering a Smart Retail Platform; 2018; 44th Euromicro Conference on Software Engineering and Advanced Applications (SEAA)

[S25] S. Latif, H. Afzaal, N. A. Zafar; Modeling of Sewerage System Using Internet of Things for Smart City; 2017; International Conference on Frontiers of Information Technology (FIT)

[S26] J. Kuklova, O. Přibyl; Framework Model in Anylogic for Smart City Ring Road Management; 2019; Smart City Symposium Prague (SCSP)

[S27] L. Walletzký, L. Carubbo, M. Ge; Modelling Service Design and Complexity for Multi-contextual Applications in Smart Cities; 2019; 23rd International Conference on System Theory, Control and Computing (ICSTCC)

[S28] E. Moustaid, S. Meijer; A hybrid approach for building models and simulations for smart cities: Expert knowledge and low dimensionality; 2017; Winter Simulation Conference (WSC)

[S29] E. Setijadi, A. K. Darmawan, R. Mardiyanto, I. Santosa, Hoiriyah, T. Kristanto; A Model for Evaluation Smart City Readiness using Structural Equation Modelling: a Citizen's Perspective; 2019; Fourth International Conference on Informatics and Computing (ICIC)

[S30] W. Jin-Xiao; Modeling for the Measurement of Smart City; 2017; International Conference on Management Science and Engineering (ICMSE)

[S31] M. Lom, O. Pribyl; Modeling Charging of Electric Vehicles in Smart Cities: Charles Square Use Case; 2019; IEEE International Conference on Connected Vehicles and Expo (ICCVE)

[S32] G. C. Lazaroiu, M. Roscia; Model for smart appliances toward smart grid into smart city; 2016; IEEE International Conference on Renewable Energy Research and Applications (ICRERA)

[S33] M. W. Nadeem, M. Hussain, M. A. Khan, M. U. Munir, S. Mehrban; Fuzzy-Based Model to Evaluate City Centric Parameters for Smart City; 2019; International Conference on Innovative Computing (ICIC)

[S34] L. Zomer, E. Moustaid, S. Meijer; A meta-model for including social behavior and data into smart city management simulations; 2015; Winter Simulation Conference (WSC)

[S35] W. Zhang, V. Terrier, X. Fei, A. Markov, S. Duncan, M. Balchanos, W. J. Sung, D. N. Mavris, M. L. Loper, E. Whitaker, M. Riley; Agent-Based Modeling of a Stadium Evacuation in a Smart City; 2018; Winter Simulation Conference (WSC)

[S36] Y. Hayashi, Y. Fujimoto, H. Ishii, Y. Takenobu, H. Kikusato, S. Yoshizawa, Y. Amano, S. Tanabe, Y. Yamaguchi, Y. Shimoda, J. Yoshinaga, M. Watanabe, S. Sasaki, T. Koike, H. Jacobsen, K. Tomsovic; Versatile Modeling Platform for Cooperative Energy Management Systems in Smart Cities; 2018; Proceedings of the IEEE

[S37] C. Mgbere, V. A. Knyshenko, A. B. Bakirova; Building Information Modeling. A Management Tool for Smart City; 2018; IEEE 13th International Scientific and Technical Conference on Computer Sciences and Information Technologies (CSIT)

[S38] A. Nuraeni, H. S. Firmansyah, G. S. Pribadi, A. Munandar, L. Herdiani, Nurwathi; Smart City Evaluation Model in Bandung, West Java, Indonesia; 2019; IEEE 13th International Conference on Telecommunication Systems, Services, and Applications (TSSA)

[S39] M. M. Rathore, Y. Jararweh, M. Raheel, A. Paul; Securing High-Velocity Data: Authentication and Key Management Model for Smart City Communication; 2019; Fourth International Conference on Fog and Mobile Edge Computing (FMEC)

[S40] Z. Xiong, Y. Zheng, C. Li; Data Vitalization's Perspective Towards Smart City: A Reference Model for Data Service Oriented Architecture; 2014; 14th IEEE/ACM International Symposium on Cluster, Cloud and Grid Computing

[S41] Giourka, P., Sanders, M.W.J.L., Angelakoglou, K., Pramangioulis, D., Nikolopoulos, N., Rakopoulos, D., Tryferidis, A., Tzovaras, D.; The smart city business model canvas—A smart city business modeling framework and practical tool; 2019; Energies

[S42] Warnecke, D., Wittstock, R., Teuteberg, F.; Benchmarking of European smart cities – a maturity model and web-based self-assessment tool; 2019; Sustainability Accounting, Management and Policy Journal

[S43] Nguyen, C., Roberts, J.J.; Green button data-access model for smart cities: Lessons learned on security, transfer, authorization, and standards-compliance in sharing energy & water usage data; 2019; Proceedings of the 2nd ACM/EIGSCC Symposium on Smart Cities and Communities, SCC

[S44]  Villegas-Ch, W., Palacios-Pacheco, X., Luján-Mora, S.; Application of a smart city model to a traditional university campus with a big data architecture: A sustainable smart campus; 2019; Sustainability (Switzerland)

[S45]  Tanda, A., De Marco, A.; Business Model Framework for Smart City Mobility Projects; 2019; IOP Conference Series: Materials Science and Engineering

[S46]  Orrego, R.B.S., Barbosa, J.L.V.; A model for resource management in smart cities based on crowdsourcing and gamification; 2019; Journal of Universal Computer Science

[S47]  Buyle, R., Van Compernolle, M., Vlassenroot, E., Vanlishout, Z., Mechant, P., Mannens, E.; Technology readiness and acceptance model as a predictor for the use intention of data standards in smart cities; 2018; Media and Communication

[S48]  Boudaa, B., Hammoudi, S., Benslimane, S.M.; Towards an Extensible Context Model for Mobile User in Smart Cities; 2018; IFIP Advances in Information and Communication Technology

[S49]  Fernández-Rodríguez, J.Y., Álvarez-García, J.A., Arias Fisteus, J., Luaces, M.R., Corcoba Magaña, V.; Benchmarking real-time vehicle data streaming models for a smart city; 2017; Information Systems

[S50]  Díaz-Díaz, R., Muñoz, L., Pérez-González, D.; Business model analysis of public services operating in the smart city ecosystem: The case of SmartSantander; 2017; Future Generation Computer Systems

[S51]  Bernardes, M.B., De Andrade, F.P., Novais, P., Lopes, N.V.; Reference model and method of evaluation for smart cities in government portals: A study of the Portuguese and Brazilian reality; 2017; Proceedings of the Internationsl Conference on Electronic Governance and Open Society: Challenges in Eurasia (eGose)

[S52]  Chen, N., Liu, Y., Li, J., Chen, Z.; A spatio-temporal enhanced metadata model for interdisciplinary instant point observations in smart cities; 2017; ISPRS International Journal of Geo-Information

[S53]  Ghosh, S., Gosavi, A.; A semi-Markov model for post-earthquake emergency response in a smart city; 2017; Control Theory and Technology

[S54]  Latorre-Biel, J.-I., Faulin, J., Jiménez, E., Juan, A.A.; Simulation model of traffic in smart cities for decision-making support: Case study in tudela (Navarre, Spain); 2017; International Conference on Smart Cities (Smart-CT)

[S55]  Ossowska, K., Szewc, L., Orłowski, C.; The principles of model building concepts which are applied to the design patterns for smart cities; 2017; 9th Asian Conference on Intelligent Information and Database Systems (ACIIDS)

[S56]  Díaz-Díaz, R., Muñoz, L., Pérez-González, D.; The Business Model Evaluation Tool for Smart Cities: Application to SmartSantander use cases; 2017; Energies

[S57]  Massana, J., Pous, C., Burgas, L., Melendez, J., Colomer, J.; Identifying services for short-term load forecasting using data driven models in a Smart City platform; 2017; Sustainable Cities and Society

[S58]  Chaturvedi, K., Kolbe, T.H.; Integrating Dynamic Data and Sensors with Semantic 3D City Models in the Context of Smart Cities; 2016; ISPRS Annals of the Photogrammetry, Remote Sensing and Spatial Information Sciences

[S59]  Li, F., Nucciarelli, A., Roden, S., Graham, G.; How smart cities transform operations models: A new research agenda for operations management in the digital economy; 2016; Production Planning and Control

[S60]  Anthopoulos, L., Fitsilis, P., Ziozias, C.; What is the source of smart city value? A business model analysis; 2016; International Journal of Electronic Government Research

[S61]  Gottschalk, M., Uslar, M.; Using a use case methodology and an architecture model for describing smart city functionalities; 2016; International Journal of Electronic Government Research

[S62]  Khari, M., Kumar, M., Vij, S., Pandey, P., Vaishali; Smart cities: A secure data transmission model; 2016; Proceedings of the Second International Conference on Information and Communication Technology for Competitive Strategies (ICTCS)

[S63]  Yu, L., Tao, S., Gao, W., Zhang, G., Lin, K.; Intelligent farm relaxation for smart city based on Internet of Things: Management system and service model; 2016; Proceedings of the International Conference on Internet of Things and Big Data (IoTBD)

[S64]  Nebiker, S., Cavegn, S., Loesch, B.; Cloud-based geospatial 3D image spaces-a powerful urban model for the smart city; 2015; ISPRS International Journal of Geo-Information

[S65]  Afonso, R.A., Dos Santos Brito, K., Do Nascimento, C.H., Garcia, V.C., Álvaro, A.; Brazilian smart cities: Using a maturity model to measure and compare inequality in cities; 2015; Proceedings of the 16th Annual International Conference on Digital Government Research (dg.o)

[S66]  Anthopoulos, L.G., Fitsilis, P.; Understanding smart city business models: A comparison; 2015; Proceedings of the 24th International Conference on World Wide Web (WWW)

[S67]  Consoli, S., Mongiovi, M., Nuzzolese, A.G., Peroni, S., Presutti, V., Recupero, D.R., Spampinato, D.; A smart city data model based on semantics best practice and principles; 2015; Proceedings of the 24th International Conference on World Wide Web (WWW)

[S68]  Bradley, P.E.; Supporting data analytics for smart cities: An overview of data models and topology; 2015; International Symposium on Statistical Learning and Data Sciences (SLDS)

[S69]  Walravens, N.; Qualitative indicators for smart city business models: The ease of mobile services and applications; 2015; Telecommunications Policy

[S70]  Perera, C., Zaslavsky, A., Christen, P., Georgakopoulos, D.; Sensing as a service model for smart cities supported by Internet of Things; 2014; Transactions on Emerging Telecommunications Technologies

[S71]  Aly C. S. Rabelo; An Architectural Model for Smart Cities using Collaborative Spatial Data Infrastructures.; 2017; SMARTGREENS

[S72]  Johannes M. Schleicher; Modeling and management of usage-aware distributed datasets for global Smart City Application Ecosystems.; 2017; PeerJ Computer Science

[S73] Esther Palomer; Component-Based Modelling for Scalable Smart City Systems Interoperability - A Case Study on Integrating Energy Demand Response Systems.; 2016; Sensors

[S74] Antonio De Nicola; A Lateral Thinking Framework for Semantic Modelling of Emergencies in Smart Cities.; 2014; International Conference on Database and Expert Systems Applications (DEXA)

[S75] Antoniundefined, Aleksandar, Marjanoviundefined, Martina, Žarko, Ivana Podnar; Modeling Aggregate Input Load of Interoperable Smart City Services; 2017; Proceedings of the 11th ACM International Conference on Distributed and Event-Based Systems

[S76] Mulligan, C.E.A., Olsson, M.; Architectural implications of smart city business models: An evolutionary perspective; 2013; IEEE Communications Magazine

[S77] Lazaroiu, G.C., Roscia, M.; Definition methodology for the smart cities model; 2012; Energy

[S78] Walravens, N.; Mobile business and the smart city: Developing a business model framework to include public design parameters for mobile city services; 2012; Journal of Theoretical and Applied Electronic Commerce Research

[S79] Sandeep Purao; Modeling Citizen-Centric Services in Smart Cities.; 2013; International Conference on Conceptual Modeling (ER)

[S80] Nils Walravens; Platform business models for smart cities - from control and value to governance and public value.; 2013; IEEE Communications Magazine

[S81] Yamagata, Y., Seya, H.; Simulating a future smart city: An integrated land use-energy model; 2013; Applied Energy

[S82] Gerhard Schmitt; Spatial modeling issues in future smart cities.; 2013; Geo-spatial Information Science

[S83] Kuk, G., Janssen, M.; The business models and information architectures of smart cities; 2011; Journal of Urban Technology

[S84] Leydesdorff, L., Deakin, M.; The triple-helix model of smart cities: A neo-evolutionary perspective; 2011; Journal of Urban Technology

[S85] Ford, M., Cadzow, S., Wilson, D., Parslow, P., Prandi, F., De Amicis, R.; Using 3D urban information models to aid simulation, analysis and visualisation of data for smart city web services (i-SCOPE); 2013; Environmental Information Systems and Services - Infrastructures and Platforms (ENVIP)

[S86] N. Walravens; Validating a Business Model Framework for Smart City Services: The Case of FixMyStreet; 2013; 27th International Conference on Advanced Information Networking and Applications Workshops

[S87] L. Zhou, Q. Li, W. Tu; An Efficient Access Model of Massive Spatiotemporal Vehicle Trajectory Data in Smart City; 2020; IEEE Access

[S88] T. K. Soltvedt, A. Sinaeepourfard, D. Ahlers; A Cost Model for Data Discovery in Large-Scale IoT Networks of Smart Cities; 2020; 21st IEEE International Conference on Mobile Data Management (MDM)

[S89] M. E. S. Cunha, R. J. F. Rossetti, P. Campos; Modelling Smart Cities Through Socio-Technical Systems; 2020; IEEE International Smart Cities Conference (ISC2)

[S90] S. C. L. Hernandes, M. E. Pellenz, A. Calsavara, M. C. Penna; A New Event Model for Event Notification Services Applied to Transport Services in Smart Cities; 2020; International Conference on Information Networking (ICOIN)

[S91] T. Alwajeeh, P. Combeau, L. Aveneau; An Efficient Ray-Tracing Based Model Dedicated to Wireless Sensor Network Simulators for Smart Cities Environments; 2020; IEEE Access

[S92] E. Grilo, B. Lopes; Modelling and Certifying Smart Cities in Reo Circuits; 2020; International Conference on Systems, Signals and Image Processing (IWSSIP)

[S93] J. Robberechts, A. Sinaeepourfard, T. Goethals, B. Volckaert; A Novel Edge-to-Cloud-as-a-Service (E2CaaS) Model for Building Software Services in Smart Cities; 2020; 21st IEEE International Conference on Mobile Data Management (MDM)

[S94] Rocha, B.D., Silva, L., Batista, T., Cavalcante, E., Gomes, P.; An Ontology-based Information Model for Multi-Domain Semantic Modeling and Analysis of Smart City Data; 2020; Proceedings of the Brazilian Symposium on Multimedia and the Web (WebMedia)

[S95] Sarv, L., Kibus, K., Soe, R.-M.; Smart city collaboration model: A case study of university-city collaboration; 2020; Proceedings of the 13th International Conference on Theory and Practice of Electronic Governance (ICEGOV)

[S96] Jovanović, D., Milovanov, S., Ruskovski, I., Govedarica, M., Sladić, D., Radulović, A., Pajić, V.; Building virtual 3D city model for smart cities applications: A case study on campus area of the university of novi sad; 2020; ISPRS International Journal of Geo-Information

[S97] Valter, P., Lindgren, P., Prasad, R.; The Future Role of Multi-business Model Innovation in a World with Digitalization and Global Connected Smart Cities; 2020; Wireless Personal Communications

[S98] Lindgren, P.; Multi Business Model Innovation in a World of Smart Cities with Future Wireless Technologies; 2020; Wireless Personal Communications

[S99] Austin, M., Delgoshaei, P., Coelho, M., Heidarinejad, M.; Architecting Smart City Digital Twins: Combined Semantic Model and Machine Learning Approach; 2020; Journal of Management in Engineering

[S100] Dos Reis, F.B., Tonkoski, R., Hansen, T.M.; Synthetic residential load models for smart city energy management simulations; 2020; IET Smart Grid

[S101] Timeus, K., Vinaixa, J., Pardo-Bosch, F.; Creating business models for smart cities: a practical framework; 2020; Public Management Review

[S102] Estrada, E., Maciel, R., Negrón, A.P.P., López, G.L., Larios, V., Ochoa, A.; Framework to support the Data Science of smart city models for decision-making oriented to the efficient dispatch of service petitions; 2020; IET Software

[S103] Sánchez-Vanegas, M.C., Dávila, M., Gutiérrez, A., López, J.D., González, L., Cobo, L., Diaz, C.O.; Towards the construction of a smart city model in Bogotá; 2020; Workshops at the Third International Conference on Applied Informatics (ICAIW)

[S104] Saeidi, A., Aghamohamadi-Bosjin, S., Rabbani, M.; An integrated model for management of hazardous waste in a smart city with a sustainable approach; 2020; Environment, Development and Sustainability

[S105] Nikoloudis, C., Strantzali, E., Tounta, T., Aravossis, K., Mavrogiannis, A., Mytilinaioy, A., Sitzimi, E., Violeti, E.; An evaluation model for smart city performance with less than 50,000 inhabitants: A Greek case study; 2020; Proceedings of the 9th International Conference on Smart Cities and Green ICT Systems (SMARTGREENS)

[S106] El-Hosseini, M., ZainEldin, H., Arafat, H., Badawy, M.; A fire detection model based on power-aware scheduling for IoT-sensors in smart cities with partial coverage; 2020; Journal of Ambient Intelligence and Humanized Computing

## REFERENCES

[1] K. M. Abbasi, T. A. Khan, and I. U. Haq. 2019. Hierarchical Modeling of Complex Internet of Things Systems Using Conceptual Modeling Approaches. *IEEE Access* 7 (2019), 102772–102791. https://doi.org/10.1109/ACCESS.2019.2930933

[2] Mohammad Abu-Matar. 2016. Towards a software defined reference architecture for smart city ecosystems. In *2016 IEEE International Smart Cities Conference (ISC2)*. 1–6.

[3] K. A. Achmad, L. E. Nugroho, A. Djunaedi, and Widyawan. 2018. Smart City Model: a Literature Review. In *2018 10th International Conference on Information Technology and Electrical Engineering (ICITEE)*. 488–493. https://doi.org/10.1109/ICITEED.2018.8534865

[4] T. Aljowder, M. Ali, and S. Kurnia. 2019. Systematic literature review of the smart city maturity model. In *International Conference on Innovation and Intelligence for Informatics, Computing, and Technologies (3ICT)*. 1–9. https://doi.org/10.1109/3ICT.2019.8910321

[5] Moussa Amrani, Dominique Blouin, Robert Heinrich, Arend Rensink, Hans Vangheluwe, and Andreas Wortmann. 2021. Multi-paradigm modelling for cyber-physical systems: a descriptive framework. *Softw. Syst. Model.* 20, 3 (2021), 611–639. https://doi.org/10.1007/s10270-021-00876-z

[6] Leonidas Anthopoulos, Marijn Janssen, and Vishanth Weerakkody. 2015. Comparing Smart Cities with Different Modeling Approaches. https://doi.org/10.1145/2740908.2743911

[7] Antoni Ballesté, Pablo Pérez-Martínez, and Agusti Solanas. 2013. The Pursuit of Citizens' Privacy: A Privacy-Aware Smart City Is Possible. *IEEE Communications Magazine* 51 (06 2013). https://doi.org/10.1109/MCOM.2013.6525606

[8] Martin Bauer, Nicola Bui, Jourik Loof, Carsten Magerkurth, Andreas Nettsträter, Julinda Stefa, and Walewski Joachim. 2013. *IoT Reference Model*. https://doi.org/10.1007/978-3-642-40403-0_7

[9] B. Benamrou, B. Mohamed, A. Bernoussi, and O. Mustapha. 2016. Ranking models of smart cities. In *4th IEEE International Colloquium on Information Science and Technology (CiSt)*. 872–879. https://doi.org/10.1109/CIST.2016.7805011

[10] Loli Burgueño, Federico Ciccozzi, Michalis Famelis, Gerti Kappel, Leen Lambers, Sébastien Mosser, Richard F. Paige, Alfonso Pierantonio, Arend Rensink, Rick Salay, Gabriele Taentzer, Antonio Vallecillo, and Manuel Wimmer. 2019. Contents for a Model-Based Software Engineering Body of Knowledge. *Softw. Syst. Model.* 18, 6 (2019), 3193–3205. https://doi.org/10.1007/s10270-019-00746-9

[11] N. Chowdhary and P. Deep Kaur. 2016. Addressing the characteristics of mobility models in IoV for smart city. In *2016 International Conference on Computing, Communication and Automation (ICCCA)*. 1298–1303. https://doi.org/10.1109/CCAA.2016.7813919

[12] Boyd Cohen. 2014. Methodology for 2014 Smart Cities Benchmarking. https://www.fastcompany.com/3038818/the-smartest-cities-in-the-world-2015-methodology

[13] C. A. de Souza, J. N. Correa, M. M. Oliveira, A. Aagaard, and M. Presser. 2019. IoT Driven Business Model Innovation and Sustainability: a literature review and a case Study in Brazil. In *2019 Global IoT Summit (GIoTS)*. 1–6. https://doi.org/10.1109/GIOTS.2019.8766371

[14] Fabian Dembski, Uwe Wössner, Mike Letzgus, Michael Ruddat, and Claudia Yamu. 2020. Urban Digital Twins for Smart Cities and Citizens: The Case Study of Herrenberg, Germany. *Sustainability* 12 (03 2020), 17p. https://doi.org/10.3390/su12062307

[15] Enspire.science. [n. d.]. TRL Scale in Horizon Europe and ERC – explained. https://enspire.science/trl-scale-horizon-europe-erc-explained/.

[16] T. T. H. Giang, M. Camargo, L. Dupont, and F. Mayer. 2017. A review of methods for modelling shared decision-making process in a smart city living lab. In *2017 International Conference on Engineering, Technology and Innovation (ICE/ITMC)*. 189–194. https://doi.org/10.1109/ICE.2017.8279888

[17] Rudolf Giffinger, Christian Fertner, H. Kramar, and Evert Meijers. 2007. City-ranking of European medium-sized cities. *Cent. Reg. Sci.* (01 2007), 1–12.

[18] Rudolf Giffinger and Haindlmaier Gudrun. 2010. Smart cities ranking: an effective instrument for the positioning of the cities? *ACE: Architecture, City and Environment* 4, 12 (2010), 7-26 pages. http://hdl.handle.net/2099/8550

[19] Ulrike Gretzel, Marianna Sigala, Zheng Xiang, and Chulmo Koo. 2015. Smart tourism: foundations and developments. *Electronic Markets* 25 (08 2015). https://doi.org/10.1007/s12525-015-0196-8

[20] A. Kontogianni, K. Kabassi, M. Virvou, and E. Alepis. 2018. Smart Tourism Through Social Network User Modeling: A Literature Review. In *2018 9th International Conference on Information, Intelligence, Systems and Applications (IISA)*. 1–4. https://doi.org/10.1109/IISA.2018.8633633

[21] Udayangani Kulatunga, Dilum Bandara, and Aravindi Samarakkody. 2020. A Systematic Review of Performance Measurement Models for Smart Cities. In *Proceedings of the International Conference on Industrial Engineering and Operations Management Dubai*.

[22] Chiehyeon Lim, Kwang-Jae Kim, and Paul Maglio. 2018. Smart cities with big data: Reference models, challenges, and considerations. *Cities* 82 (05 2018). https://doi.org/10.1016/j.cities.2018.04.011

[23] Meiyi Ma, Sarah M. Preum, Mohsin Y. Ahmed, William Tärneberg, Abdeltawab Hendawi, and John A. Stankovic. 2019. Data Sets, Modeling, and Decision Making in Smart Cities: A Survey. *ACM Trans. Cyber-Phys. Syst.* 4, 2, Article 14 (Nov. 2019), 28 pages. https://doi.org/10.1145/3355283

[24] Jeremy Millard, Rasmus Kåre Thaarup, Jimmy Kevin Pederson, Catriona Manville, Matthias Wissner, Bas Kotterink, Gavin Cochrane, Jonathan Cave, Andrea Liebe, and Roel Massink. 2014. Mapping smart cities in the EU. *Direzione generale delle Politiche interne dell'Unione (Parlamento europeo)*

(2014). https://doi.org/10.2861/3408

[25] M. Mishbah, B. Purwandari, and D. I. Sensuse. 2018. Systematic Review and Meta-Analysis of Proposed Smart Village Conceptual Model: Objectives, Strategies, Dimensions, and Foundations. In *2018 International Conference on Information Technology Systems and Innovation (ICITSI)*. 127–133. https://doi.org/10.1109/ICITSI.2018.8696029

[26] Vaia Moustaka, Athena Vakali, and Leonidas Anthopoulos. 2018. A Systematic Review for Smart City Data Analytics. *Comput. Surveys* 51 (12 2018), 1–41. https://doi.org/10.1145/3239566

[27] Gabriela Pereira, Peter Parycek, Enzo Falco, and Reinout Kleinhans. 2018. Smart governance in the context of smart cities: A literature review. *Information Polity* 23 (05 2018), 1–20. https://doi.org/10.3233/IP-170067

[28] Charith Perera, Yongrui Qin, Julio C. Estrella, Stephan Reiff-Marganiec, and Athanasios V. Vasilakos. 2017. Fog Computing for Sustainable Smart Cities: A Survey. *ACM Comput. Surv.* 50, 3, Article 32 (June 2017), 43 pages. https://doi.org/10.1145/3057266

[29] Kai Petersen, Robert Feldt, Shahid Mujtaba, and Michael Mattsson. 2008. Systematic Mapping Studies in Software Engineering. *Proceedings of the 12th International Conference on Evaluation and Assessment in Software Engineering* (06 2008).

[30] Manuel Pedro Rodríguez Bolívar. 2018. Fostering the Citizen Participation Models for Public Value Creation in Cooperative Environment of Smart Cities. In *Electronic Government*. Springer International Publishing, Cham, 235–248.

[31] Umberto Rosati and Sergio Conti. 2016. What is a Smart City Project? An Urban Model or A Corporate Business Plan? *Procedia - Social and Behavioral Sciences* 223 (2016), 968 – 973. https://doi.org/10.1016/j.sbspro.2016.05.332

[32] Karen Rose, Scott Eldridge, and Lyman Chapin. 2015. The internet of things: An overview. *The internet society (ISOC)* 80 (2015), 1–50.

[33] Francisca Rosique, Fernando Losilla, and Juan Angel Pastor. 2018. A Domain Specific Language for Smart Cities. In *4th International Electronic Conference on Sensors and Applications*.

[34] Maria Teresa Rossi et al. 2023. Replication package of the work titled "A Systematic Mapping Study on Smart Cities Modeling Approaches". https://drive.google.com/drive/folders/151n74-F47FEAcJmxm43nxt5J4LqwnHIn?usp=sharing.

[35] Davide Di Ruscio, Dimitrios S. Kolovos, Juan de Lara, Alfonso Pierantonio, Massimo Tisi, and Manuel Wimmer. 2022. Low-code development and model-driven engineering: Two sides of the same coin? *Softw. Syst. Model.* 21, 2 (2022), 437–446.

[36] Apurvanand Sahay, Arsene Indamutsa, Davide Di Ruscio, and Alfonso Pierantonio. 2020. Supporting the understanding and comparison of low-code development platforms. In *46th Euromicro Conference on Software Engineering and Advanced Applications (SEAA)*. 171–178. https://doi.org/10.1109/SEAA51224.2020.00036

[37] Ruben Sanchez-Corcuera, Adrián Núñez-Marcos, Jesus Sesma-Solance, Aritz Bilbao, Rubén Mulero Martínez, Unai Zulaika, Gorka Azkune, and Aitor Almeida. 2019. Smart cities survey: Technologies, application domains and challenges for the cities of the future. *International Journal of Distributed Sensor Networks* 15 (2019). https://doi.org/10.1177/1550147719853984

[38] Mary Shaw. 2002. What makes good research in software engineering? *Int. J. Softw. Tools Technol. Transf.* 4, 1 (2002), 1–7. https://doi.org/10.1007/s10009-002-0083-4

[39] N. Shetty, S. Renukappa, S. Suresh, and K. Algahtani. 2019. Smart City Business Models – A Systematic Literature Review. In *2019 3rd International Conference on Smart Grid and Smart Cities (ICSGSC)*. 22–25. https://doi.org/10.1109/ICSGSC.2019.00-24

[40] Dorota Sikora-Fernandez and Danuta Stawasz. 2016. THE CONCEPT OF SMART CITY IN THE THEORY AND PRACTICE OF URBAN DEVELOPMENT MANAGEMENT. *Romanian Journal of Regional Science* 10 (06 2016), 86–99.

[41] G. Solmaz and D. Turgut. 2019. A Survey of Human Mobility Models. *IEEE Access* 7 (2019), 125711–125731. https://doi.org/10.1109/ACCESS.2019.2939203

[42] P. Torrinha and R. J. Machado. 2017. Assessment of maturity models for smart cities supported by maturity model design principles. In *2017 IEEE International Conference on Smart Grid and Smart Cities (ICSGSC)*. 252–256. https://doi.org/10.1109/ICSGSC.2017.8038586

[43] Hans Vangheluwe and Juan de Lara. 2003. Foundations of multi-paradigm modeling and simulation: computer automated multi-paradigm modelling: meta-modelling and graph transformation. In *Proceedings of the 35th Winter Simulation Conference: Driving Innovation*. 595–603. https://doi.org/10.1109/WSC.2003.1261474

[44] Roel Wieringa, Neil Maiden, Nancy Mead, and Colette Rolland. 2005. Requirements Engineering Paper Classification and Evaluation Criteria: A Proposal and a Discussion. *Requir. Eng.* 11, 1 (2005), 102–107. https://doi.org/10.1007/s00766-005-0021-6

[45] Claes Wohlin, Per Runeson, Martin Hst, Magnus C. Ohlsson, Bjrn Regnell, and Anders Wessln. 2012. Experimentation in Software Engineering.

[46] Andreas Wortmann, Olivier Barais, Benoît Combemale, and Manuel Wimmer. 2020. Modeling languages in Industry 4.0: an extended systematic mapping study. *Softw. Syst. Model.* 19, 1 (2020), 67–94. https://doi.org/10.1007/s10270-019-00757-6